\documentclass[prx,twocolumn,superscriptaddress,nofootinbib,showpacs,notitlepage]{revtex4-2}
\usepackage{amsmath}
\usepackage{amsfonts}
\usepackage{amssymb}
\usepackage{bbm}
\usepackage[utf8]{inputenc}
\usepackage{braket}
\usepackage{environ}
\usepackage{graphicx,color,dsfont}
\usepackage{booktabs}
\usepackage{floatrow}
\usepackage[caption=false,label font={bf,normalsize}]{subfig}
\usepackage[dvipsnames]{xcolor}
\usepackage[all]{xy}
\usepackage{relsize}
\usepackage{hyperref}
\usepackage{accents}
\hypersetup{linktocpage}
\definecolor{darkblue}{RGB}{0,0,127} 
\definecolor{darkgreen}{RGB}{0,150,0}
\hypersetup{colorlinks=true,citecolor=darkblue,linkcolor=darkblue, 
	filecolor=red, urlcolor=darkblue,pdftitle={Fault-tolerant quantum computation on non-orientable manifold: application to topological stabilizer and Floquet codes},
pdfauthor={Ryohei Kobayashi, Guanyu Zhu}}
\usepackage{lmodern}

\usepackage{lipsum}

\makeatletter
\def\@opargbegintheorem#1#2#3{\trivlist
   \item[]{\bfseries #1\ #2\ (#3)} \itshape}
\makeatother

\NewEnviron{eqs}{%
\begin{equation}\begin{split}
    \BODY
\end{split}\end{equation}
}

\newcommand{\ZZ}{\mathbb{Z}}
\newcommand{\Z}{\mathbb{Z}}
\DeclareMathOperator*{\Motimes}{\text{\raisebox{0.25ex}{\scalebox{0.8}{$\bigotimes$}}}}

\begin{document}

\title{Cross-cap defects and fault-tolerant logical gates in the surface code and the honeycomb Floquet code}

\author{Ryohei Kobayashi}
\email{ryok@umd.edu}
\affiliation{Department of Physics, Condensed Matter Theory Center, and Joint Quantum Institute, University of Maryland, College Park, Maryland 20742, USA}

\author{Guanyu Zhu}
\email{guanyu.zhu@ibm.com}
\affiliation{IBM Quantum, IBM T.J. Watson Research Center, Yorktown Heights, NY 10598 USA}
 
\begin{abstract}
We consider the $\mathbb{Z}_2$ toric code, surface code and Floquet code defined on a non-orientable surface, which can be considered as families of codes extending Shor's 9-qubit code. We investigate the fault-tolerant logical gates of the $\Z_2$ toric code in this setup, which corresponds to $e\leftrightarrow m$ exchanging symmetry of the underlying $\Z_2$ gauge theory.
 We find that non-orientable geometry provides a new way the emergent symmetry acts on the code space, and discover the new realization of the fault-tolerant Hadamard gate of 2d surface code with a single cross-cap connecting the vertices non-locally along a slit, dubbed a non-orientable surface code. This Hadamard gate can be realized by a constant-depth local unitary circuit modulo non-locality caused by a cross-cap. 
Via folding, the non-orientable surface code can be turned into a bi-layer local quantum code, where the folded cross-cap is equivalent to a bi-layer twist terminated on a gapped boundary and the logical Hadamard only contains local gates with intra-layer couplings when being away from the cross-cap, as opposed to the inter-layer couplings on each site needed in the case of the folded surface code.  We further obtain the complete logical Clifford gate set for a stack of non-orientable surface codes and similarly for codes defined on Klein-bottle geometries.
We then construct the honeycomb Floquet code in the presence of a single cross-cap, and find that the period of the sequential Pauli measurements acts as a $HZ$ logical gate on the single logical qubit, where the cross-cap enriches the dynamics compared with the orientable case. We find that the dynamics of the honeycomb Floquet code is precisely described by a condensation operator of the $\Z_2$ gauge theory, and illustrate the exotic dynamics of our code in terms of a condensation operator supported at a non-orientable surface.

\end{abstract}

\maketitle

\tableofcontents

\section{Introduction}
Quantum error-correcting codes stand as a cornerstone of fault-tolerant quantum computation. In the past few decades, there has been active effort of finding new error-correcting codes and associated logical gates, due to their importance from both theoretical and practical perspectives.
In many cases, a quantum error-correcting code realizes a topologically ordered state at its code space, where the robust nature of topological order against local perturbations makes it possible to achieve fault-tolerance~\cite{Kitaev1997, Dennis:2002ds}. 
One of the most familiar examples is the $\Z_2$ toric code that realizes the $\Z_2$ gauge theory at its ground state subspace, and many well-known codes such as Shor’s 9-qubit code, the Steane code, and the Reed-Muller code are regarded as storing the topologically ordered states supported on a non-trivial spatial manifold at their code spaces~\cite{freedman1998projective, campbell2017roads}.

The topologically ordered states can generally be realized in two distinct ways; 
one approach is the traditional ``passive'' way, which prepares the state as a ground state of a specific many-body Hamiltonian. The other is the ``active'' way that prepares and maintains a state by sequentially measuring local operators, instead of considering a fixed Hamiltonian. The widely studied codes in this category are the stabilizers codes, where commuting stabilizers can be simultaneously measured in each round.
As another remarkable example, Hastings and Haah recently discovered a protocol to realize the topologically ordered state by a periodic sequence of non-commuting two-qubit Pauli measurements, dubbed a honeycomb Floquet code~\cite{Hastings2021honeycomb}.
Here, the code implements an instantaneous stabilizer group with the sequence of measurements, which stabilizes a topologically ordered state equivalent to the $\Z_2$ toric code.
Interestingly, a period of the measurement schedule on the honeycomb Floquet code induces an emergent $\Z_2$ symmetry for the $e\leftrightarrow m$ exchange of anyons in the $\Z_2$ toric code, which leads to a non-trivial logical gate acting on the code space. The study of Floquet code and their new variants is under active development in recent years, including the Floquet code with boundaries~\cite{Haah2022boundaries}, with dead qubits~\cite{aasen2023dead}, with insertion of defects~\cite{ellison2023floquet}, automorphism codes~\cite{Aasen2022automorphism, davydova2023automorphism}, 
CSS version~\cite{davydova2023CSS}, and generalization to three space dimensions~\cite{dua2023engineering}. 

For the purpose of fault-tolerant quantum computation, one needs to implement fault-tolerant logical gates which are automorphisms $U$ acting on the code spaces~$\mathcal{C}$, i.e., $U:\mathcal{C} \rightarrow \mathcal{C}$. A very important class of such logical gates correspond to  \textit{transversal gates} which can be expressed as a product of unitary $\Motimes_j U_j$, or more generally \textit{constant-depth (geometrically) local circuits} \cite{Eastin:2009cj, Bravyi:2013dx}.  While transversal gates do not propagate pre-existing errors within each code block, constant-depth local circuits have very limited error propagation constrained by a constant-size light cone \cite{Bravyi:2013dx}.   In the context of topological codes, such transversal gates or constant-depth logical circuits $U$ correspond to the \textit{emergent global symmetries} of the underlying topological order or topological quantum field theory (TQFT) \cite{Zhu:2017tr, Yoshida_gate_SPT_2015, Yoshida16, Yoshida17, Beverland:2016bi, Webster_gates_2018, zhu:2022fractal, Barkeshli2023codim2, barkeshli2023highergroup}, or sometimes also referred to as topological or anyon symmetries in the case of (2+1)D topological order \cite{barkeshli2014SDG}.  In general, the unitary $U$ is called an emergent symmetry when it keeps the low-energy Hilbert space of a system invariant. In this specific case of topological codes, the emergent symmetry  $U$ keeps the ground-state subspace $\mathcal{C}$ of the corresponding topological order (equivalent to the code space) invariant.    More specifically, transversal gates correspond to onsite symmetries of the underlying topological order \cite{Zhu:2017tr}, while general constant-depth circuits may not correspond to onsite symmetries.  We also note that in general  $U$ does not have to keep the underlying system Hamiltonian $H$ invariant (i.e., $UHU^\dag =H$), which is only required in the case when $U$ is also an exact microscopic symmetry.    While many existing transversal logical gates are equivalent to exact onsite microscopic symmetries,  there can be a much larger variety of logical gates only corresponding to emergent symmetries, which are not necessarily transversal but can be realized by more general constant-depth local circuits in the stabilizer codes or sequence of measurements in the Floquet codes as will be demonstrated explicitly in this paper. 

Another related class of logical gates are the so-called \textit{fold-transversal gates}, which were first introduced in the context of identifying a triangular color code with a folded surface codes \cite{Kubica:2015br, Moussa:2016} and then generalized to generic topological codes including non-Abelian topological codes \cite{Zhu:2017tr} (also related to modular transformations) as well as to quantum low-density parity-check (LDPC) codes \cite{breuckmann2022fold}.  One can view such fold-transversal gates as transversal gates on a folded code, which contains inter-layer gates applied transversally on the folded geometry.  One can also understand the fold-transversal gate from the perspective of emergent symmetries: the folding turns geometrically non-local reflection symmetry in the underlying topological order into local onsite symmetry in a bi-layer topological order with the folded edge corresponding to a particular gapped boundary, as pointed out in Ref.~\cite{Zhu:2017tr}.  In the case of surface code, the logical Hadamard gate can be implemented via a fold-transversal gate with the combination of transversal Hadamard and pairwise SWAPs between reflection-symmetric points with respect to the diagonal folding axis, which is turned into pairwise inter-layer SWAPs in the folded surface code.  Similarly, the logical phase ($S$) gate can also be implemented via the fold-transversal gate, and together with the transversal logical CNOT gate and the logical Hadamard gate form a complete logical Clifford gate set. 

Nevertheless, since the fold-transversal gate is, strictly speaking, a constant-depth circuit but not a constant-depth local circuit in the unfolded geometry due to the long-range gates coupling reflection-symmetric points, the error propagation is hence not subjected to a constant-size light cone. There is hence an obvious disadvantage for the fold-transversal gate:  the coupling between pair of qubits acted by the long-range gates will lead to long-range correlated noise, which may potentially lead to an absence of fault-tolerant threshold, which has been addressed in Ref.~\cite{Bluvstein_2023, cain2024correlated}. However, it remains an open question whether it is possible to obtain the same type of logical Clifford gates on a $\ZZ_2$ toric code without introducing a significant reduction of the effective code distance. 

In this paper, we unlock a new realization of fault-tolerant logical gates on the topological stabilizer and Floquet codes by putting a code on a \textit{non-orientable} surface. It turns out that the non-orientable geometry allows us to enrich the way that the emergent global symmetry of the topological order acts on the Hilbert space, which naturally leads to the intriguing class of logical gates. 

The study of quantum codes and topologically ordered states on non-orientable geometry has a rather long history. Indeed, the well-known Shor's 9-qubit code~\cite{shor1995} can be thought of as the $\Z_2$ toric code defined on the real projective plane $\mathbb{RP}^2$ equipped with a specific triangulation.
The $\Z_2$ toric code and the double semion model on $\mathbb{RP}^2$ are studied in Ref.~\cite{freedman1998projective, Freedman2016doublesemion} respectively. The cross-cap geometry introduced to make up a non-orientable surface can be thought of inserting a defect of the spatial reflection symmetry; from field theoretical perspective, the response of reflection or time-reversal symmetry on topological phases was extensively studied by studying non-orientable spacetime manifold in e.g., Ref.~\cite{Chan2016nonorientable, Witten:2016cio, Barkeshli2019reflection, Shapourian2017, Tata2022anomalies}.

Our main contribution is the discovery of new fault-tolerant logical gates of the error-correcting codes enabled by employing non-orientable geometry, for both topological stabilizer and Floquet code. As we will describe below, it in particular allows us to implement the fault-tolerant logical Hadamard gate and also the whole logical Clifford gate set on the $\Z_2$ toric code without the introduction of long-range correlated noise, as well as find enriched dynamics of the Floquet code driven by the measurement schedule carried out on the non-orientable surface.   The constant-depth logical gate considered in this paper is particularly relevant in a 3D pancake architecture \cite{Landahl:2011vs} consisting of a stack of 2D topological codes where the transversal or constant-depth logical gates can be applied in a single shot on each individual code block or between the neighboring code blocks.  These single-shot logical gates are especially important when including an additional 3D toric or color codes to implement non-Clifford CCZ or T gates and than code-switch to 2D topological codes to implement the whole Clifford logical gate set \cite{Bombin:2015jk, Vasmer2019, Kubica:2021}.

\subsection{Summary of results}
Here we provide a brief summary of our main results.
After reviewing the implementation of the logical gate in a standard surface code in Sec.~\ref{sec:overview}, we start with constructing the $\Z_2$ toric code on a 2d torus with a single cross-cap in Sec.~\ref{sec:RP2code}, which we call a non-orientable toric code.
The cross-cap is introduced in the geometry by considering a thin slit in the torus, and then connecting the vertices at the antipodal points along the slit by edges, which makes the resulting geometry non-orientable.
A notable property of this setup is that cross-cap geometry can store a single logical qubit; the $e,m$ string operators with support at the homologically equivalent loop passing through the cross-cap have the non-trivial commutation relation, and give a pair of the logical Pauli operators $\{\overline{Z},\overline{X}\}$ acting a single logical qubit. 
Thanks to the fact that the logical Pauli operators $\overline{Z},\overline{X}$ are supported at the homologically same loop, one can find a realization of the logical Hadamard gate $\overline{H}: \overline{Z}\leftrightarrow \overline{X}$ by a constant-depth local circuit, generating the $e\leftrightarrow m$ exchanging symmetry in the presence of a single cross-cap. In Sec.~\ref{subsec:hadamardRP2}, we explicitly construct this fault-tolerant Hadamard gate of the non-orientable toric code using a local constant-depth circuit which causes the stabilizers to flow to the nearby location. Note that this local constant-depth circuit corresponds to an emergent symmetry only preserving the code space, rather than an exact microscopic symmetry preserving the parent Hamiltonian as in the case of the folded surface code.  

When folding the system along the reflection line cutting through the cross-cap, one can realize the cross-cap geometry by (geometrically) local stabilizers and hence obtain a (geometrically) \textit{local quantum code} \cite{portnoy2023local}. In particular, the folded cross-cap is composed of stabilizers with local inter-layer couplings, which is topologically equivalent to a \textit{bi-layer twist} corresponding to layer-exchanging $\Z_2$ symmetry (also called \textit{genon}) \cite{barkeshli2013genon} terminated at a gapped boundary equivalent to the fold.  The logical Hadamard gate is still a constant-depth local circuit in this folded code.  One can also further move the bi-layer twist away from the gapped boundary corresponding to the fold, which results in a Klein-bottle geometry while the logical Hadamard gate can still be implemented via the same constant-depth local circuit if treating the additional logical qubit as an ancilla to be traced out.

Armed with the realization of the logical Hadamard gate, in Sec.~\ref{subsec:clifford} we provide the realization of the logical Clifford gate set for a stack of non-orientable surface codes, which are surface codes with a single cross-cap. In particular, one can then realize the logical phase ($S$) gate with a fold-transversal gate similar to the case of the folded surface code, and the standard transversal logical CNOT gates between two copies of non-orientable surface codes which are equivalent to CSS codes.  

In Sec.~\ref{subsec:kleinbottle}, we also construct a translational-symmetric realization of the  $\Z_2$ toric code on a Klein bottle, where the logical Hadamard gate can be realized by the combination of transversal Hadamard and a single lattice translation of the code. The lattice translation can also be implemented by a constant-depth local circuit, mediated by additional auxiliary physical qubits introduced to realize the lattice translation by a shallow sequence of local SWAP operators.  Such a constant-depth local circuit turns out to implement the exact microscopic symmetry of the parent Hamiltonian, in contrast to the emergent symmetry in the case of the non-orientable surface code.   Nevertheless, since the cross-cap in this code does not have a thin-slit geometry, there is no way to fold the code on a Klein bottle into a local quantum code, which reflects the flexibility of emergent symmetries compared to the exact microscopic symmetries.

Compared with the logical Hadamard gate in the case of the single-layer surface code which needs $O(d^2)$ long-range connections,   the non-orientable surface code only needs $O(d)$ static long-range connections to engineer the cross-cap.    For the folded geometries, the logical Hadamard in the folded surface code needs inter-layer SWAP while only intra-layer coupling is needed in the folded non-orientable surface code when being away from the cross-cap.  Since long-range or inter-layer couplers/gates typically have lower fidelity than short-range intra-layer couplers/gates in most experimental platforms, it is expected that the logical gate in the non-orientable surface codes will have higher fidelity.  As opposed to the logical Hadamard ($H$) and phase ($S$) gate on the surface code implemented via the fold-transversal gates, the error propagation caused by the logical Hadamard gate in the non-orientable surface code does not suffer from long-range correlated noise due to its local constant-depth nature, as shown in Sec.~\ref{subsec:errorpropagation}. In addition, for large connected errors, the constant-depth circuit gate only propagate errors by $O(1)$ distance near the boundary of the error cluster, in contrast to a factor of two increase of error support in the case of fold-transversal gates.
On the other hand, the logical phase ($S$) gate in the non-orientable surface code is still implemented via a fold-transversal gate which suffers from the same problem of error-propagation and long-range correlated noise as in the case of the folded surface code.  Nevertheless, one can use the lower-fidelity logical $S$ gate to prepare a logical $Y$ state in an ancilla logical qubit and use higher fidelity logical $H$ and CNOT to implement a logical $S$ gate on the target logical qubit. One can then use several rounds of state distillation protocol to gradually increase the fidelity of the logical $Y$ state and the logical $S$ gate.

In Sec.~\ref{sec:floquet}, we construct the honeycomb Floquet code on a 2d torus with a single cross-cap, where we can perform the sequence of two-qubit Pauli measurement in the presence of a single cross-cap. The cross-cap again stores a single logical qubit protected by the instantaneous stabilizer group. 
One can see that a period of the measurements acts by the $\overline{H}\overline{Z}$ logical gate on the single logical qubit for the cross-cap, realizing the $e\leftrightarrow m$ exchanging symmetry. Interestingly, the presence of the cross-cap enriches the dynamics of the honeycomb Floquet code; one period of the measurement schedule $\overline{H}\overline{Z}$ for the $e\leftrightarrow m$ exchange generates the $\Z_4$ group, while it generates the $\Z_2$ symmetry in the orientable case. 

The above properties of the honeycomb Floquet code with a cross-cap can be derived by expressing one period of the measurement schedule by a ``condensation operator'' for the fermionic particle $\psi$ as described in Sec.~\ref{subsec:condensation}; namely, one can express the action of one period of measurements by the sum of the string operators for the $\psi$ particle over all possible configurations of closed loops. As described in Ref.~\cite{seifnashri2022condensation}, this sum over the $\psi$ closed loops generates the emergent symmetry exchanging the anyons $e\leftrightarrow m$ in the $\Z_2$ gauge theory. The expression of the dynamics in the honeycomb Floquet code in terms of the condensation operator gives an additional insight for deriving its action on the code space, and useful for understanding the case of the non-orientable surface, including the enlarged period of dynamics from $\Z_2$ to $\Z_4$.

 We conclude this paper by discussions of the possible generalizations in Sec.~\ref{sec:discussions}, including the expectation about the fault-tolerant logical $T$ gate on (4+1)D $\Z_2$ toric code with an emergent fermionic particle. In Appendix~\ref{sec:kleinmeasurement}, we also describe the realization of the logical Hadamard gate on Klein bottle by Pauli measurements, on the Wen plaquette-translation code recently proposed in~\cite{aasen2023measurement}.

\section{Overview: logical gate of $\Z_2$ toric code}
\label{sec:overview}
In this section, we briefly review known results about the logical gate of the (2+1)D $\Z_2$ toric code, and also describe several simplest realizations of the (2+1)D $\Z_2$ toric code on a non-orientable manifold to illustrate the essential idea of this paper.

\subsection{Review: surface code and its logical Hadamard gate}
\label{subsec:surface}
\begin{figure*}
    \centering
    \includegraphics[width=1\textwidth]{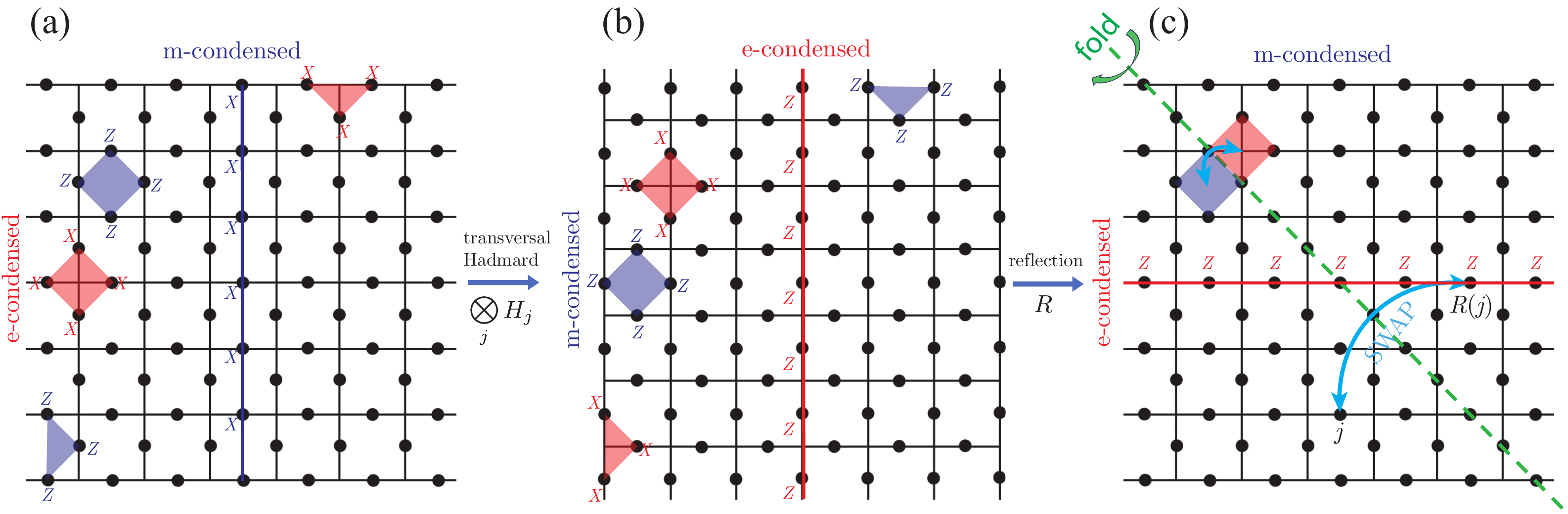}
\caption{Logical Hadamard gate of the surface code. (a): The boundary condition of the surface code is realized by the $e$-condensed on the vertical boundary and $m$-condensed on the horizontal boundary. (b): By applying the transversal Hamadard gate, the bulk stabilizers are switched, and the labels for the boundary condition is switched $e\leftrightarrow m$ as well. (c): By further applying the SWAP gate for each pair of qubits related by the reflection along the diagonal line, one can bring the code back to the original surface code.}
\label{fig:surface-code}
\end{figure*}

First, let us recall the implementation of the logical Hadamard gate on the well-known \textit{surface code}; the $\Z_2$ toric code on a planar surface with $e$-condensed and $m$-condensed boundaries. Fig.~\ref{fig:surface-code}(a) shows a surface code with four-body bulk $X$-stabilizers ($Z$-stabilizers) on the vertices (plaquettes) and three-body boundary $X$-stabilizers ($Z$-stabilizers).  The $m$-condensed boundary ($e$-condensed boundary), composed of boundary $X$-stabilizers, condenses the $m$ particles ($e$ particles) which hence allows logical $\overline{X}$ ($\overline{Z}$) operator, i.e., a string of Pauli $X$ ($Z$), to terminate on the boundary.  

The logical Hadamard gate of the surface code is associated with the exact microscopic symmetry of the code, which also corresponds to a \textit{fold-transversal gate} \cite{Kubica:2015br, Moussa:2016, Zhu:2017tr, breuckmann2022fold}.

To describe this $\Z_2$ symmetry of the surface code, we first consider applying a transversal Hadamard operation $\Motimes_j H_j$ on all the qubits, as illustrated in Fig.~\ref{fig:surface-code}(b). Due to the transformation $H_j : X_j \leftrightarrow Z_j$,  all the $X$ and $Z$  stabilizer operators switch, as illustrated via the exchange of vertices and plaquettes.  
This transversal Hadamard gate obviously does not respect the symmetry of the surface code, since it exchanges $X$ (vertex) and $Z$ (plaquette) stabilizers as well as the boundary conditions ($e$-condensed $\leftrightarrow$ $m$-condensed) as illustrated in Fig.~\ref{fig:surface-code}(a,b), which hence maps the original code space $\mathcal{C}$ to a different code space $\mathcal{C}'$. 
One can obtain the exact symmetry transformation within the original code space $\mathcal{C}$ by further applying a  reflection $R$ about the diagonal axis  following the transversal Hadamard. This spatial reflection can be implemented by pairwise SWAPs between all  reflection-symmetric qubits denoted by $\text{SWAP}(j \leftrightarrow R\cdot j)$, as illustrated in Fig.~\ref{fig:surface-code}(c). Note that the reflection exchanges the vertices and plaquettes of the lattice.
The combined transformation implements the logical Hadamard gate $\overline{H}$ since it acts on the logical Pauli operators as $\overline{X}\leftrightarrow \overline{Z}$, while mapping the code space $\mathcal{C}$ back to itself. It is also called a fold-transversal gate \cite{Kubica:2015br, Moussa:2016, Zhu:2017tr, breuckmann2022fold}. At the level of effective $\Z_2$ gauge theory, this symmetry is understood as the combination of the $e\leftrightarrow m$ exchanging symmetry together with the spatial reflection symmetry, i.e., 
\begin{equation}\label{eq:Hadamard_surface}
\overline{H} = R U_{em} \equiv \text{SWAP}\big(j \leftrightarrow R(j)\big) \Motimes_j H_j. 
\end{equation}
Since this combined transformation $R U_{em}$ preserves the parent Hamiltonian of the surface code, it is an exact global symmetry.

To ease the practical implementation, one can also turn this spatial symmetry into an onsite symmetry by folding the surface code along the diagonal axis \cite{Moussa:2016}, as illustrated in Fig.~\ref{fig:surface-code}(c).  The same logical Hadamard gate now corresponds to the transversal Hadamard combined with the layer swap: 
\begin{equation}
\overline{H} = \Motimes_j ( H_{j}^{(1)} H_{j}^{(2)} \text{SWAP}_{j}^{(1 \leftrightarrow 2)} ). 
\end{equation}

\subsection{$\Z_2$ toric code on a non-orientable surface: simple examples}
\label{subsec:unorientedZ2tc}
One of the main purposes of this paper is to achieve the fault-tolerant logical gates of the $\Z_2$ toric code on a non-orientable manifold. Before moving to the explicit construction of the logical gates, let us illustrate such non-orientable codes in simplest setups.
In Fig.~\ref{fig:non_orientable_toric_code_simple}, we present several possible realizations of non-orientable geometry in the $\Z_2$ toric code. Below let us explain each of them in order.

\begin{figure*}
    \centering
    \includegraphics[width=0.8\textwidth]{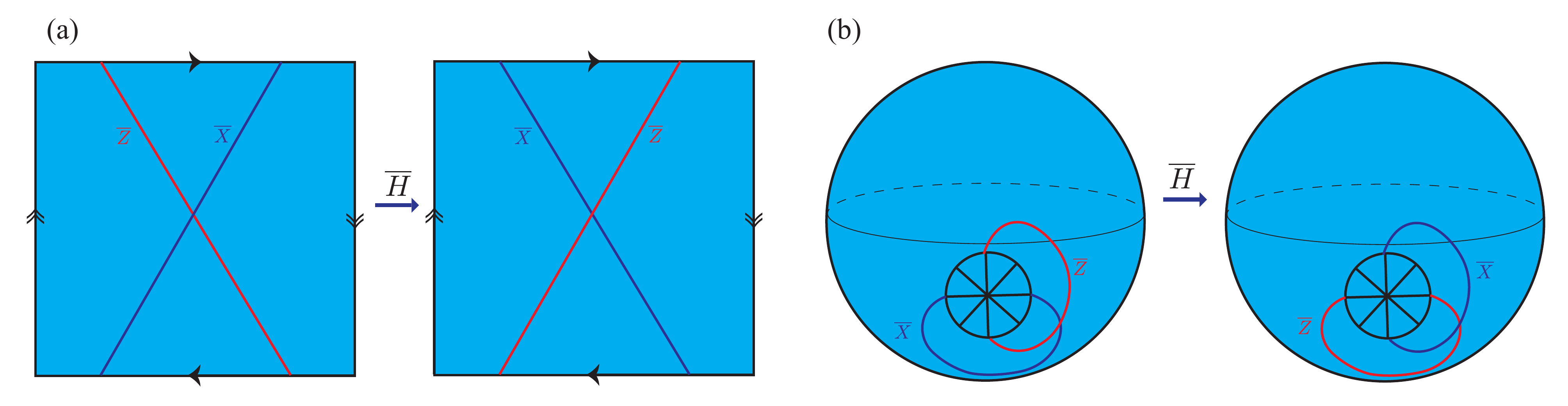}
\caption{Schematic figure for the action of the logical Hadamard gate on the Pauli logical gates of the non-orientable toric code. (a): $\mathbb{RP}^2$ is represented by a square whose boundary condition identifies the antipodal points on the boundary. On $\mathbb{RP}^2$, the logical $\overline{Z}$, $\overline{X}$ gate of $\Z_2$ toric code is both supported at the single non-contractible loop. The logical Hadamard gate exchanges these two logical gates. (b): This figure describes essentially the same setup as (a), but $\mathbb{RP}^2$ is represented by a sphere with a single cross-cap.}
\label{fig:RP2_TQFT}
\end{figure*}

First, we consider in Fig.~\ref{fig:non_orientable_toric_code_simple} (a) a square-patch toric code with inverted periodic boundary condition corresponding to $\mathbb{RP}^2$, which is a lattice realization of the geometry shown in Fig.~\ref{fig:RP2_TQFT}(a).  The circles on the vertices represent qubits. The top and bottom circles with the same label (1,2,3) are identified and hence represent the same qubit.  Similarly, the qubits on the left and right with the same label (I, II, III) are also identified.
The white and black plaquettes represent 4-body $X$ stabilizers  $S^X_p $$=$$ \Motimes_{v\subset \partial p} X_v$ and $Z$ stabilizers $S^Z_{p'} $$=$$ \Motimes_{v\subset \partial p} Z_v$ respectively.   Here, $p$ and $p'$ label the white and black plaquettes respectively.

Second, we implement in Fig.~\ref{fig:non_orientable_toric_code_simple}(b) a surface with a cross-cap in the center, which is a microscopic realization of the geometry in Fig.~\ref{fig:RP2_TQFT}(b).   In particular, the cross-cap is implemented by connecting vertices related by a reflection symmetry on a thin slit.   This gives rise to geometrically non-local stabilizers as illustrated in Fig.~\ref{fig:non_orientable_toric_code_simple}(b), the majority of which are 4-body stabilizers occupying a pair of neighboring vertices and their inverted counterparts with respect to the central point.   There is a single 8-body non-local $Z$-stabilizers located on the two ends of the slit. 

Finally, we recall in Fig.~\ref{fig:non_orientable_toric_code_simple}(c) that the well-known Shor's 9-digit code also belongs to the family of the $\Z_2$ toric code on a non-orientable surface $\mathbb{RP}^2$~\cite{freedman1998projective}; Fig.~\ref{fig:non_orientable_toric_code_simple}(c) shows a specific cellulation of $\mathbb{RP}^2$ that contains nine edges and three vertices. We put a qubit on each edge, so the $\Z_2$ toric code is formed by 9 physical qubits. There are three $S^X$ stabilizers defined on each vertex, $S^X_v=\Motimes_{v\subset \partial e} X_e$ only two of which are independent, and six $S^Z$ stabilizers $S^X_p=\Motimes_{e\subset \partial p} X_e$ involving two edges among the three connecting a fixed pair of vertices. The above set of eight $S^X, S^Z$ stabilizers precisely realizes the original Shor's 9-digit code.

\subsection{Logical gate and emergent symmetry}
In general, a logical gate of the topological code is understood as an emergent symmetry of the system, as a unitary leaving the low-energy Hilbert space invariant.
In particular, when the emergent symmetry is generated by a local constant-depth unitary circuit, the corresponding logical gate achieves fault-tolerance since it spreads the error only locally.
See Refs.~\cite{Zhu:2017tr, Yoshida_gate_SPT_2015, Yoshida16, Yoshida17, Beverland:2016bi, Webster_gates_2018, zhu:2022fractal, Barkeshli2023codim2, barkeshli2023highergroup} for detailed discussions about the connection between invertible symmetry of topological codes and fault-tolerant logical gates.

\subsubsection{Hadamard gate of $\Z_2$ toric code as an emergent symmetry}
Let us discuss the potential fault-tolerant logical gate of the family of the $\Z_2$ toric code on a non-orientable surface, based on the emergent symmetry of the effective $\Z_2$ gauge theory.
For simplicity, we mainly focus on the real projective plane $\mathbb{RP}^2$, which is regarded as a sphere with a single cross-cap.   The code space $\mathcal{C}$ of the $\Z_2$ toric code on $\mathbb{RP}^2$ is given by
\begin{equation}
    \mathcal{C} = \mathbb{C}^{|H_1(\mathbb{RP}^2,  \mathbb{Z}_2)|} = \mathbb{C}^2,
\end{equation}
where $H_1(\mathbb{RP}^2,  \mathbb{Z}_2) = \mathbb{Z}_2$ represents the first $\ZZ_2$ homology group.  The fact that the code space is two-dimensional (encoding one logical qubit) corresponds to the fact that there is only a single non-trivial homology class $[\gamma_1]$$\in$$H_1(\mathbb{RP}^2, \mathbb{Z}_2)$ on the real projective plane.   In contrast to the surface code or the toric code on an orientable surface, the logical $\overline{Z}$ and $\overline{X}$ string operators are supported on the same fundamental cycle $\gamma_1$.

At the level of the effective $\Z_2$ gauge theory, 
the $e\leftrightarrow m$ exchanging symmetry of the $\Z_2$ gauge theory is identified as the logical Hadamard gate; the logical gate $\overline{X}$ corresponds to the line operator for the $m$ particle on the non-contractible cycle, and transformed by the $e\leftrightarrow m$ exchanging symmetry to $\overline{Z}$ for the $e$ particle supported on the same cycle, which has the same action as the logical Hadamard gate (see Fig.~\ref{fig:RP2_TQFT}). Hence, one can expect that the logical Hadamard gate on $\mathbb{RP}^2$ can be achieved solely by the local constant-depth unitary circuit generating the $e\leftrightarrow m$ exchanging symmetry. Indeed, in Sec.~\ref{sec:RP2code} we will explicitly construct the Hadamard gate via local constant-depth circuit on the specific stabilizer code on a non-orientable surface. 

Note that the realization of the logical Hadamard gate by a local constant-depth circuit is achievable on the $\Z_2$ toric code only if we put the system on a non-orientable surface; otherwise one cannot find the logical $\overline X$ and $\overline Z$ operators supported on the same fundamental cycle, so they cannot be transformed into each other by the local constant-depth operations. Indeed, as we have seen previously, the surface code also requires the non-local spatial reflection symmetry to obtain the logical Hadamard gate, and we need to fold the geometry to make the operation local.

\subsubsection{Logical gates implemented by sweeping invertible defects}
Before closing this section, let us briefly recall the relationship between the logical gates and the invertible defects of the stabilizer code, following~\cite{Yoshida16}.
For simplicity, we restrict ourselves to the case of a codimension-1 defect of the (2+1)D stabilizer code, which corresponds to 0-form emergent symmetry acting on the whole 2d space.

\begin{figure}
    \centering \includegraphics[width=0.5\textwidth]{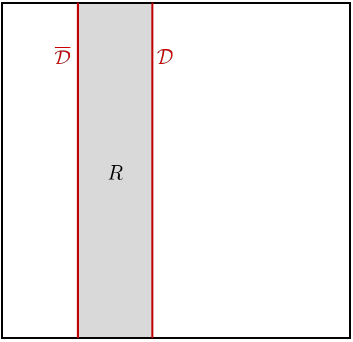}
\caption{Codimension-1 defect of the emergent symmetry is obtained by applying the symmetry on the restricted region $R$ in the 2d space. The logical gate is understood as sweeping a defect over the whole space.}
\label{fig:defect}
\end{figure}

As we mentioned earlier, a fault-tolerant logical gate of the stabilizer code is understood as an emergent symmetry of the parent Hamiltonian generated by a local constant-depth unitary circuit. For a given fault-tolerant logical gate, one can construct a defect of the emergent symmetry by acting the symmetry generator on a restricted region $R$ in the 2d space; it amounts to modifying the Hamiltonian at the boundary of the region $\partial R$, which is understood as an insertion of a codimension-1 symmetry defect $\mathcal{D}$ at $\partial R$. By choosing the region $R$ to be a thin slab, one can create a pair of defects $\mathcal{D}, \overline{\mathcal{D}}$ which can be annihilated by fusing them together (see Fig.~\ref{fig:defect}). In this sense, the defect is said to be invertible.
Conversely, the logical gate can be understood by ``sweeping'' the invertible defect $\mathcal{D}$ over the whole 2d space by expanding the region $R$ where the circuit for the emergent symmetry is applied.

\begin{figure*}
    \centering
    \includegraphics[width=0.8\textwidth]{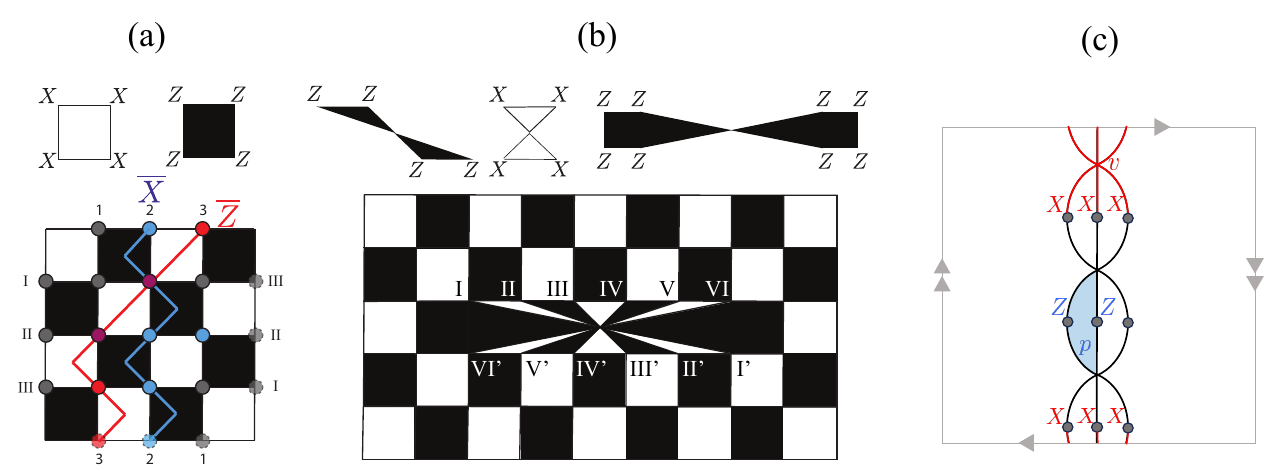}
\caption{Possible realizations of the $\Z_2$ toric code on a non-orientable surface. (a): The boundary condition of the surface code on a square is implemented so that the vertices at the antipodal points on the boundary are identified. This realizes the cross-cap at the boundary, and gives a topoloogy equivalent to a real projective plane $\mathbb{RP}^2$. (b): Another way to implement a cross-cap is to insert a slit in the bulk of $\Z_2$ toric code (which looks like $1\times l$ rectangle with $l$ the length of the slit), and connect the a pair of vertices at the antipodal points along the slit by an edge, i.e., connect the vertices labeled by the same number.
(c): The Shor's 9-qubit code is known to be regarded as a $\Z_2$ toric code on $\mathbb{RP}^2$. The figure shows the graph on which the $\Z_2$ toric code is constructed, where the $X$ and $Z$ stabilizers of the $\Z_2$ toric code together define a code identical to the Shor's 9-digit code.
}
\label{fig:non_orientable_toric_code_simple}
\end{figure*}

\subsection{The pancake architecture: a 3D architecture with a stack of 2D codes }
Since this paper focuses on fault-tolerant logical gates with constant-depth local circuits or transversal-like gates, we consider a particular experimental architecture where such logical gates become relevant.
This is the so-called ``pancake" architecture of fault-tolerant quantum computation \cite{Landahl:2011vs}, where one stacks 2D code blocks into a 3D architecture. We note a multi-layer architecture is achievable for the superconducting qubit platform \cite{Bravyi:2024wc}.   In this architecture,  transversal logical gates or constant-depth local circuits can be applied to these codes to perform fault-tolerant quantum computation.  In particular, for any pair of identical CSS codes, it is well-known that the transversal CNOT gate between the two code blocks gives rise to a logical CNOT gate.  This allows transversal logical CNOT between neighboring code blocks in the pancake architecture.   One good candidate code for this architecture is the 2D color code, since one can apply all logical Clifford gates transversally on 2D triangular color codes \cite{Bombin:2006hw, Landahl:2011vs}. Since a 2D triangular color code is equivalent to a folded surface code \cite{Kubica:2015br, Moussa:2016}, one can alternatively stack folded surface codes into a pancake architecture \cite{Bravyi:2020vw}, which has a larger error threshold due to the smaller stabilizer weight compared to the color code.   

Due to the Eastin-Knill no-go theorem \cite{Eastin:2009cj}, one needs additional operation besides transversal gates to make the computation universal. One can either use magic state distillation protocol \cite{bravyi2005} or modify the pancake architecture with an additional 3D color code which can be used to perform non-Clifford gate followed by coding switch to a 2D color code \cite{Bombin:2015jk}.
In fact, reference~\cite{Kubica:2021} has done a systematic comparison of these two approaches, and find the later one performs better when the physical error rate is low enough while the first one is better in the higher error rate regime.   We note that in a usual 2D surface code architecture, the logical Clifford gates can be implemented via lattice surgery protocols which needs $O(d)$ time overhead to ensure fault tolerance \cite{Brown:2016td}.  On the other hand, the single-shot transversal logical Clifford gate in the 3D pancake architecture has time overhead only $O(1)$, therefore has advantage over the 2D architecture even when choosing the magic-state distillation approach to achieve universality.   Besides achieving universality, the 3D pancake architecture is also useful to demonstrate a fault-tolerant version of quantum advantage of constant depth circuit \cite{Bravyi:2020vw}.  

In this paper, we mainly focus on improving the single-shot  logical Clifford gates in the context of a stack of surface codes forming a 3D pancake architecture. It can be potentially applied to the scenario where one performs all logical Clifford gates in 2D surface code and then code switch to a a 3D toric code or color code to perform a logical non-Clifford CCZ or T gates in  \cite{Vasmer2019, Bombin:2015jk}. Alternatively, the non-Clifford gates can also be done with magic-state distillation which requires single-shot logical Clifford gates as elementary building blocks.  In addition, the techniques developed in this work can also be potentially modified to improve the fold-transversal gates in quantum LDPC codes. Relevant hardware architecture can be a multi-layer superconducting qubit architecture.

\section{Construction of non-orientable toric code and its logical gates}
\label{sec:RP2code}
Here we explicitly describe the realization of the logical Hadamard gate for a single qubit by a local constant-depth circuit, in the microscopic model of the $\Z_2$ toric code on a non-orientable surface. As outlined in the previous section, this is performed by implementing $e\leftrightarrow m$ symmetry of the $\Z_2$ toric code in the presence of a single cross-cap.

\subsection{Warm-up: $e\leftrightarrow m$ exchange of $\Z_2$ toric code by constant-depth circuit}
\begin{figure*}
    \centering
    \includegraphics[width=0.7\textwidth]{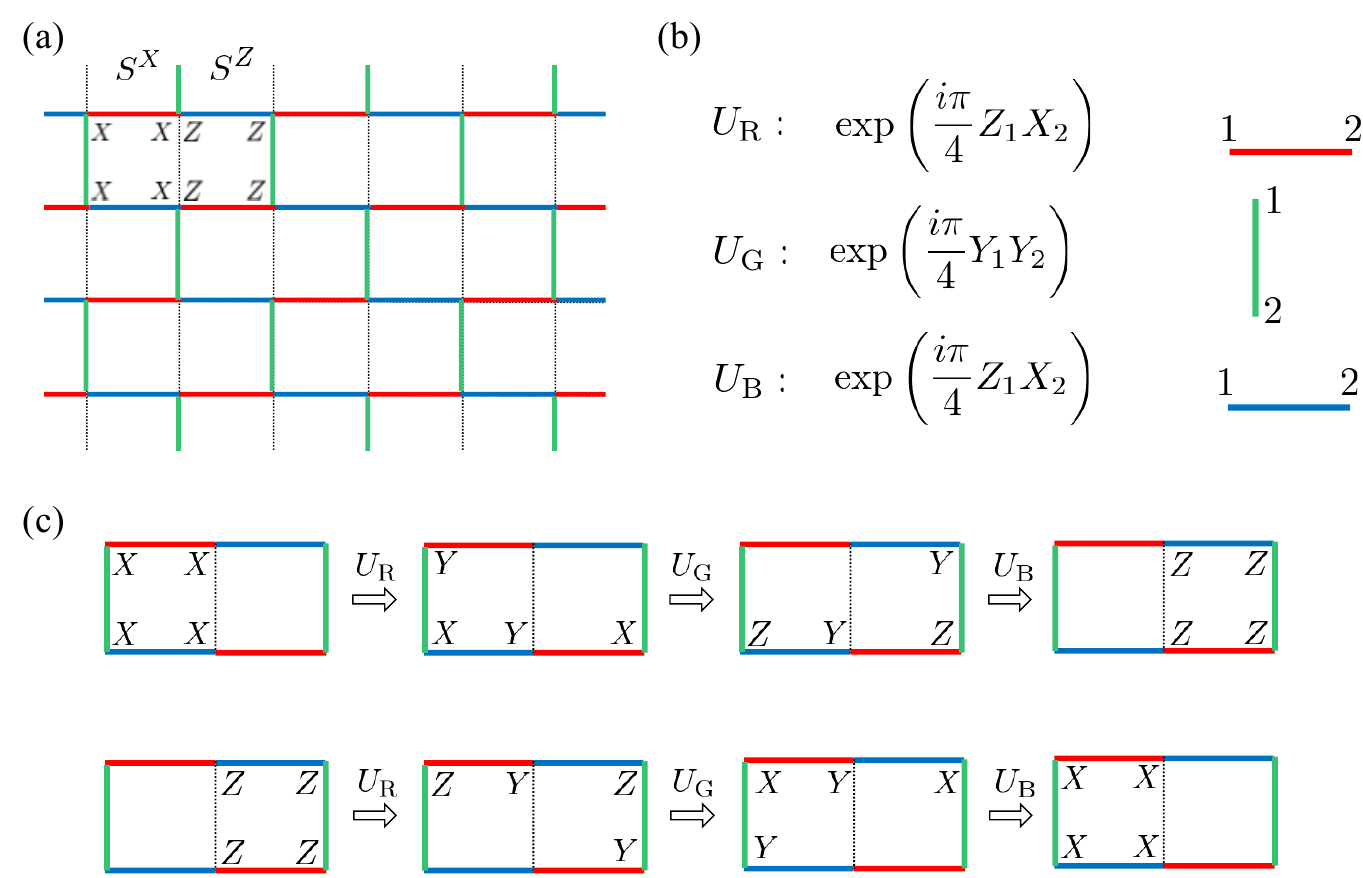}
\caption{$e\leftrightarrow m$ exchanging symmetry of the $\Z_2$ toric code by a unitary $U=U_{\mathrm{B}}U_{\mathrm{G}}U_{\mathrm{R}}$. (a): we assign the red, green, and blue coloring on the edges of the square lattice. (b): Each unitary $U_{\mathrm{R}}$, $U_{\mathrm{G}}$, $U_{\mathrm{B}}$ is defined as the product of local unitaries supported on the edges with the corresponding color. (c): The unitary $U$ acts on the stabilizers $S^X, S^Z$ by permuting them on the pair of neighboring plaquettes. }
\label{fig:torichadamard}
\end{figure*}

Before we explicitly describe the Hadamard gate on a non-orientable geometry, we illustrate the realization of $e\leftrightarrow m$ exchanging symmetry of $\Z_2$ toric code on an orientable surface in terms of a local constant-depth circuit, following Ref.~\cite{Barkeshli2023codim2}.
We consider a square-patch toric code with the plaquette stabilizers $S^X$ and $S^Z$, see Fig.~\ref{fig:torichadamard} (a).
The logical gate is expressed by a sequence of three unitaries as $U= U_{\mathrm{B}}U_{\mathrm{G}}U_{\mathrm{R}}$, where the definition of each unitary is given in Fig.~\ref{fig:torichadamard} (a),(b).
While one can also realize the $e\leftrightarrow m$ symmetry by a single lattice translation of the toric code associated with the transversal Hadamard gate, 
note that this unitary $U$ does not induce lattice translation. This unitary $U$ rather induces the local permutation of the two stabilizers $S^X$ and $S^Z$ within a pair of neighboring plaquettes, see Fig.~\ref{fig:torichadamard} (c).
This internal nature of the unitary $U$ is desirable for finding $e\leftrightarrow m$ exchanging gate in the presence of a single cross-cap, since one cannot define a lattice translation symmetry compatible with a single cross-cap.  

\subsection{Non-orientable toric code and its Hadamard gate}
\label{subsec:hadamardRP2}
Now we describe a specific realization $\Z_2$ toric code in the presence of a single cross-cap, on which we will construct the logical Hadamard gate. 

\begin{figure*}
    \centering
    \includegraphics[width=1\textwidth]{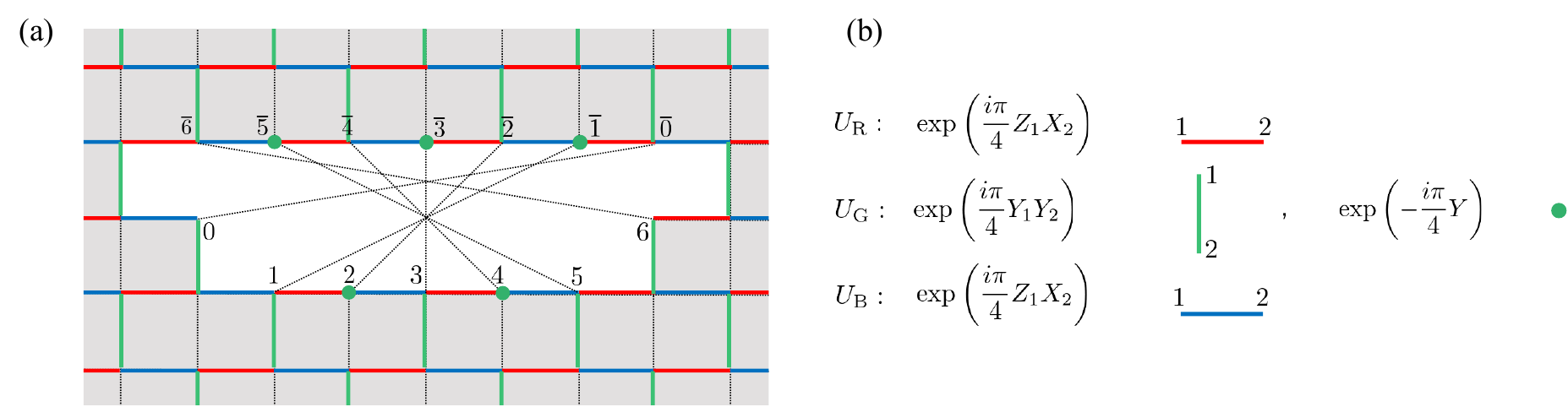}
\caption{The cross-cap introduced in the $\Z_2$ toric code and the $e\leftrightarrow m$ exchange of the code. (a): we connect the vertices $j$, $\overline{j}$ by an edge, which makes up a square lattice with a cross-cap. (b): the definition of the unitary $U=U_{\mathrm{B}}U_{\mathrm{G}}U_{\mathrm{R}}$ in the case with a cross-cap. Note that the definition of $U_{\mathrm{G}}$ has an additional unitary at the cross-cap compared with the oriented case, where its acts by $\exp(-\frac{i\pi}{4}Y)$ on a single qubit represented by a green dot.}
\label{fig:crosscaphadamard}
\end{figure*}

We define a code on a 2d torus with a single cross-cap. 
Due to a technical reason, we consider a realization of a cross-cap as described in Fig.~\ref{fig:crosscaphadamard} (a), which is different from simplest realizations shown in Fig.~\ref{fig:non_orientable_toric_code_simple}. 
We label the vertices at the cross-cap by numbers $0\le j \le 2m$, and $\overline{j}$ represents the antipodal vertex of $j$. Fig.~\ref{fig:crosscaphadamard} (a) depicts the case where $m=3$.

\begin{figure*}
    \centering
    \includegraphics[width=0.7\textwidth]{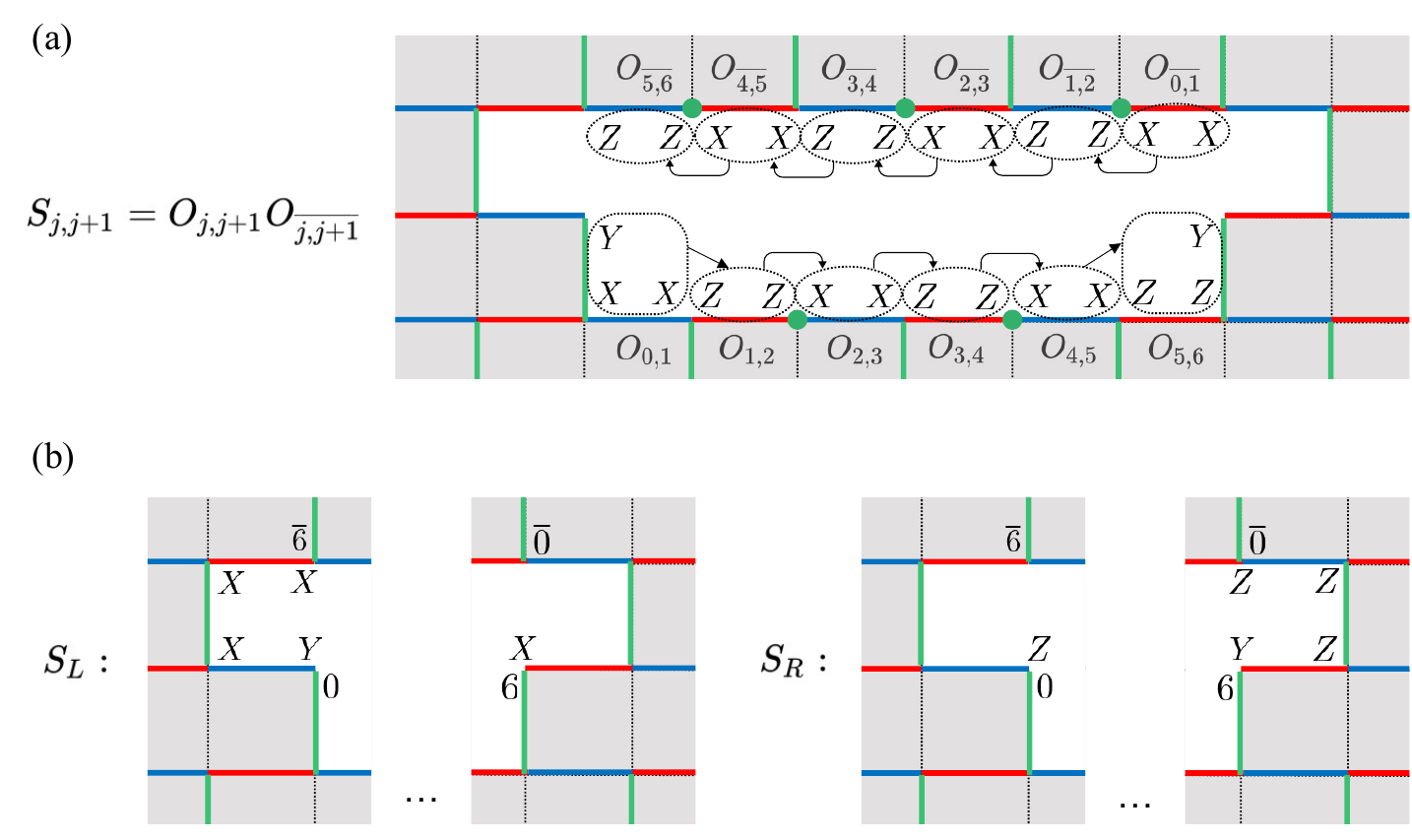}
\caption{The stabilizers near the cross-cap. (a): the stabilizer $S_{j,j+1}$ for $0\le j\le 2m-1$ is defined as $S_{j,j+1} = O_{j,j+1}O_{\overline{j,j+1}}$. The arrows in the figure mean that the unitary $U$ transforms $O_{j-1,j}\to O_{j,j+1}, O_{\overline{j-1,j}}\to O_{\overline{j,j+1}}$ for $1\le j\le 2m-1$. (b): the additional two stabilizers $S_L,S_R$ at the end of the cross-cap, both of which act on five qubits.}
\label{fig:stabilizercrosscap}
\end{figure*}

Let us define the stabilizer group. For the square plaquettes not on the cross-cap region, we have the stabilizer of the $\Z_2$ toric code as usual, i.e., $S^X$ and $S^Z$. 
To describe the stabilizers at the cross-cap, we define the operators $O_{j,j+1},O_{\overline{j,j+1}}$ with $0\le j \le 2m-1$, which are described in Fig.~\ref{fig:stabilizercrosscap} (a) for the case of $m=3$. We then define the stabilizers as
\begin{align}
    S_{j,j+1} = O_{j,j+1}O_{\overline{j,j+1}},
\end{align}
i.e., we combine the operators $O$'s located at the antipodal points of the cross-cap.
We also define an additional two stabilizers located at the end of the cross-cap, $S_L$ and $S_R$, described in Fig.~\ref{fig:stabilizercrosscap} (b). This completes the whole set of generators of our stabilizer group. For simplicity, this code will be called the ``non-orientable toric code'' in this paper.

The code distance $d$ is given by the length of the shortest path going through the cross-cap exactly once, so has the order $d=\Theta(m)$.~\footnote{Here $f=\Theta(g)$ means $\lim\sup_{n\to\infty}f/g <\infty$ and $\lim\inf_{n\to\infty}f/g >0$, while $O(n)$ only corresponds to the former condition.}

The local constant-depth circuit $U=U_{\mathrm{B}}U_{\mathrm{G}}U_{\mathrm{R}}$ can again be defined for the non-orientable toric code as well, as shown in Fig.~\ref{fig:crosscaphadamard} (b). Note that
the definition of $U_{\mathrm{G}}$ has an additional unitary at the cross-cap represented as a green dot in Fig.~\ref{fig:crosscaphadamard} that acts as $\exp(-\frac{i\pi}{4}Y)$ on a single qubit at the green vertex.

Let us check that the unitary $U$ works as the logical gate. For the square plaquettes away from the cross-cap, it acts as the permutation of the plaquette stabilizers $S^Z\leftrightarrow S^X$ according to Fig.~\ref{fig:torichadamard} (c).

For the operators on the cross-cap, one can check that  $O_{j,j+1},O_{\overline{j,j+1}}$ for $1\le j\le 2m-1$ get ``translated'' along the cross-cap, as shown in Fig.~\ref{fig:stabilizercrosscap}. This implies 
\begin{align}
    S_{j-1,j}\to S_{j,j+1}\quad \text{for $1\le j\le 2m-1$.}
\end{align}
Also, the unitary $U$ acts on the stabilizers near the end of the cross-cap as
\begin{align}
S_L\to S_R,\quad S_R\to S_{0,1}S_{R}, \quad S_{2m-1,2m}\to S_LS_R.
\end{align}
One can then see that the whole stabilizer group is preserved by the action of $U$, hence a logical gate. Note that since $U$ does not commute with the Hamiltonian near the cross-cap but preserves the code space, $U$ now generates an emergent symmetry of the non-orientable toric code, while it was exact in the absence of a cross-cap.
By studying the action of $U$ on the $X, Z$ string operators passing through the cross-cap once, one can explicitly check that $U$ induces the action on the logical Pauli operators implemented by a cross-cap as $\overline{X}\to\overline{Z}, \overline{Z}\to -\overline{X}$. This action corresponds to the logical gate $\exp(-\frac{i\pi}{4}\overline{Y}) = \overline{H}\overline{Z}$, so it realizes the Hadamard gate $\overline{H}$ up to the logical Pauli $\overline{Z}$ operator.

Note that since we are considering a 2d torus with a cross-cap, there are two logical qubits stored by cycles of the torus, apart from a single logical qubit stored by a cross-cap. 
The logical gates of the non-orientable toric code presented in this paper always act on the code space by the form of $U_{\mathrm{torus}}\otimes U_{\mathrm{cc}}$, where $U_{\mathrm{torus}}$, $ U_{\mathrm{cc}}$ act within the code space for the torus and the cross-cap respectively.
Therefore, one can simply regard the two logical qubits for the torus as an ancilla, and identify the unitary $U$ as the $\overline{H}\overline{Z}$ gate by simply focusing on the single logical qubit while discarding the state for the torus. 

Also, note that the non-orientable toric code is equivalent to a CSS code by the local unitary $\exp(\frac{i\pi}{4}Z_{0})\exp(-\frac{i\pi}{4}X_{2m})$ support at two vertices $0,2m$, which can properly transform the Pauli $Y$ operators appearing in the stabilizers $S_{0,1},S_{2m-1,2m},S_L,S_R$ into $X$ or $Z$. As we will see in the later subsection, this CSS property enables us to construct complete Clifford gate set for the copies of the codes.

\subsection{Realizing a local non-orientable toric code on a bilayer system via folding}
\label{subsec:folding}

\begin{figure*}
    \centering
    \includegraphics[width=1\textwidth]{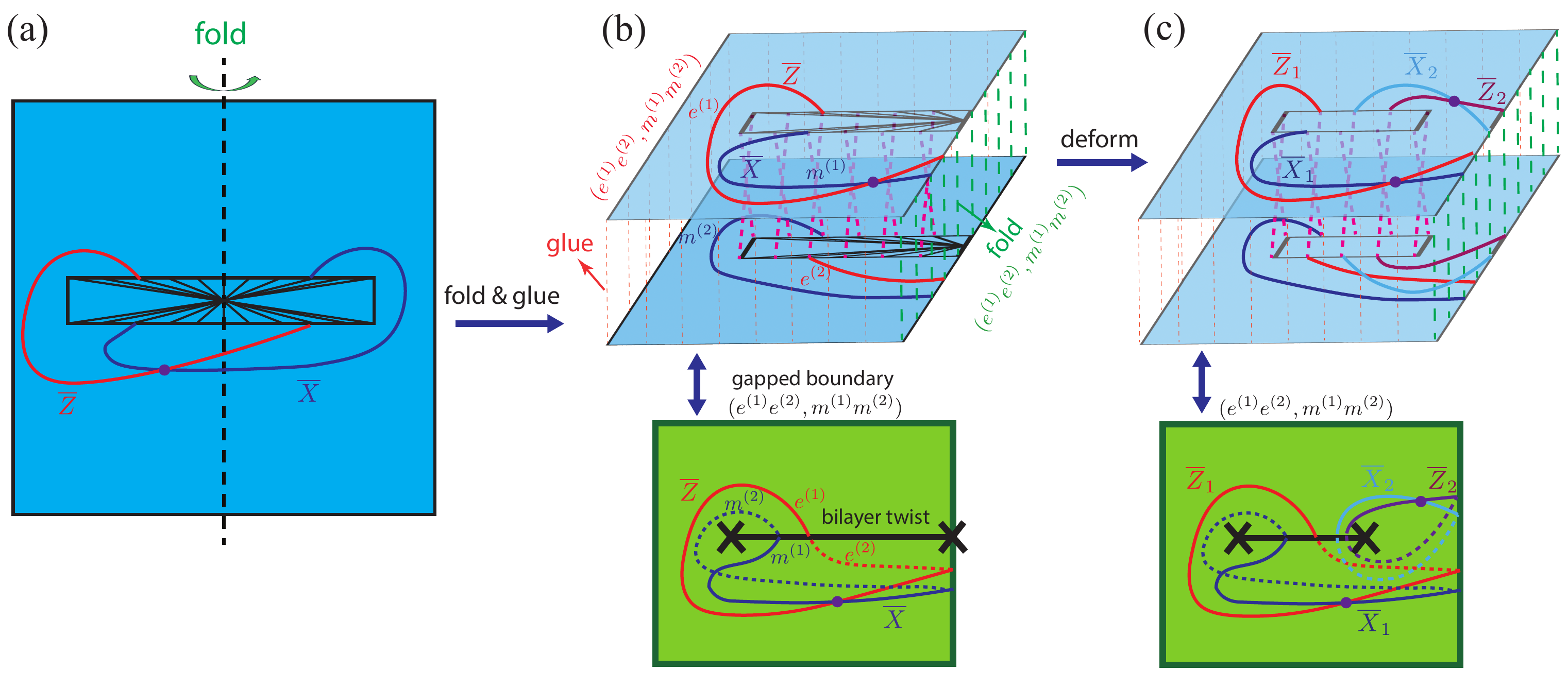}
\caption{(a) The non-orientable toric code is folded into a bi-layer geometry along the axis cutting through the cross-cap.  (b) The long-range coupling of the cross-cap is turned into local inter-layer coupling in the folded non-orientable toric code as shown in the upper panel, which is equivalent to a bilayer twist (genon) terminated on the fold.  The other three edges on both layers are also glued together to form a sphere with a  cross-cap, i.e., the $\mathbb{RP}^2$ geometry.  All edges on the folded non-orientable toric code including the fold and the glued edges correspond to the gapped boundary of the type $(e^{(1)}e^{(2)}, m^{(1)}m^{(2)})$.  (c) One can deform the folded setup by moving the bilayer twist away from from the fold, which leads to a Klein-bottle geometry storing two logical qubits, where the corresponding logical operators are shown explicitly.  }
\label{fig:folding_bilayer}
\end{figure*}

\begin{figure*}
    \centering
    \includegraphics[width=1\textwidth]{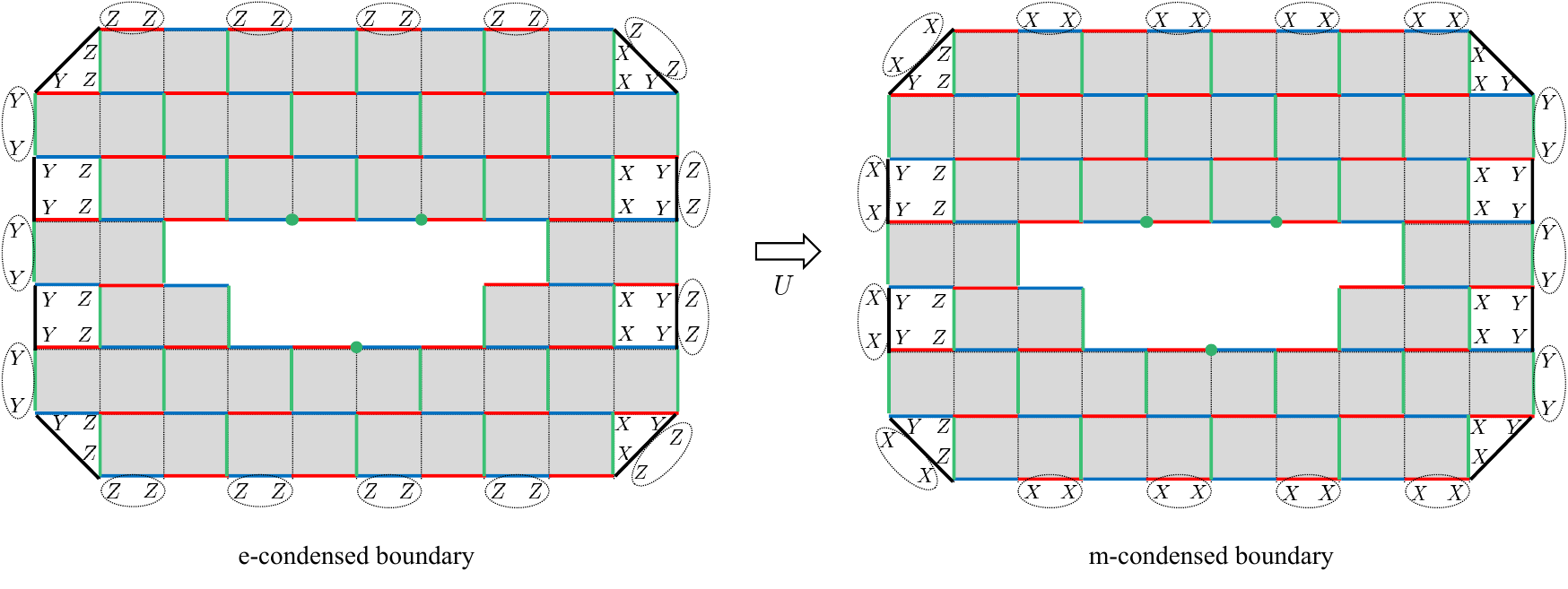}
\caption{The gapped boundary conditions of the $\Z_2$ toric code and the action of the unitary $U = U_{\mathrm{B}}U_{\mathrm{G}}U_{\mathrm{R}}$ on the boundary condition. 
First, one can see that there are white plaquettes (white triangle or square) near the boundary; the stabilizers on these white plaquettes are invariant under the action of $U$. They are not regarded as the boundary stabilizers condensing the anyons, but rather as a part of bulk stabilizers of the $\Z_2$ toric code.
Then, the left figure represents the $e$-condensed boundary condition, where the boundary stabilizer condensing $e$ is represented by the operators encircled. By the action of $U$, each boundary stabilizer flows clockwise to the boundary stabilizers in the right figure, up to multiplication by a stabilizer on a white plaquette. So, $U$ acts by a logical operation which transforms the left stabilizer group to the right one. The right figure represents the $m$-condensed boundary.}
\label{fig:boundarytrans}
\end{figure*}

\begin{figure*}
    \centering   \includegraphics[width=1\textwidth]{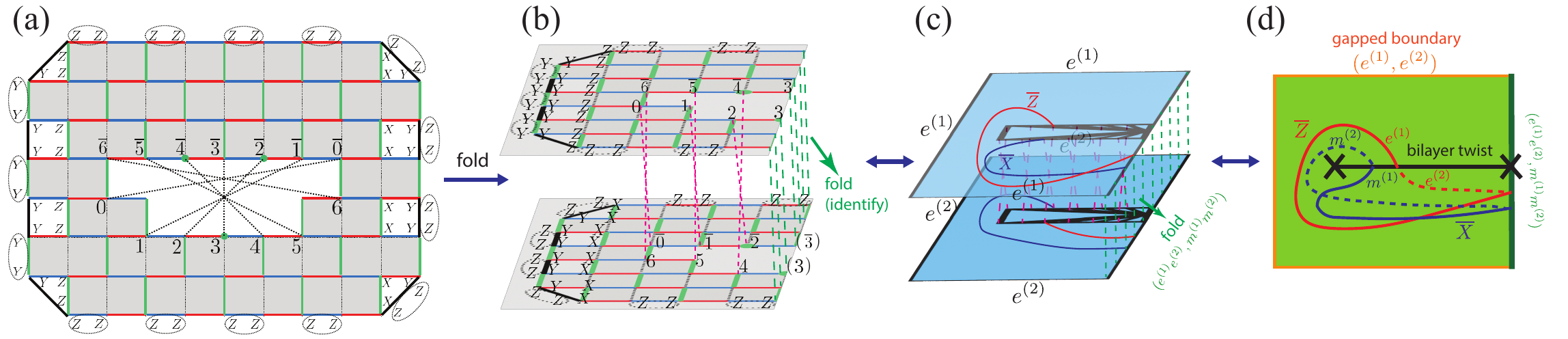}
\caption{(a) The non-orientable surface code with $e$-condensed external boundary.  (b,c,d) The folded non-orientable surface code with $e$-condensed external boundaries on both layers, while the fold corresponds to the $(e^{(1)}e^{(2)}, m^{(1)}m^{(2)})$-boundary. }
\label{fig:folded_lattice}
\end{figure*}


\subsubsection{Folding the non-orientable toric code and layer-exchanging defect}
The non-orientable toric code can only be implemented on a 2d torus with long-range connectivity since the stabilizer across the cross-cap involves qubits separated by large spatial distance, as can be seen in Fig.~\ref{fig:non_orientable_toric_code_simple} and Fig.~\ref{fig:crosscaphadamard}.    For practical implementation of the code, one should consider turning it into a \textit{local quantum code}, i.e., a code defined on a lattice with geometric locality \cite{portnoy2023local}. When the code is located on a square with open boundary and a cross-cap  (see Fig.~\ref{fig:folding_bilayer} (a) for a location of the cross-cap), the solution is to fold the code along the vertical line cutting the cross-cap to make the cross-cap geometry local and glue the open boundaries together. This makes it into a bi-layer system with geometric locality which is topologically equivalent to an  $\mathbb{RP}^2$ manifold, i.e., a sphere with a cross-cap.   In order to achieve local connectivity after folding, the cross-cap needs to have a slit-like geometry as in Fig.~\ref{fig:non_orientable_toric_code_simple} and Fig.~\ref{fig:crosscaphadamard}. Folding the cross-cap into a layered system with locality is illustrated in Fig.~\ref{fig:folding_bilayer}(a,b).

In the folded code, we get a new type of geometrically local symmetry defect, i.e., the folded cross-cap (denoted by $c$) terminated on the fold (green-dashed lines), as illustrated in Fig.~\ref{fig:folding_bilayer}(b). One can actually consider this folded cross-cap as a defect of the layer-exchanging $\Z_2$ symmetry $\rho_{S_{12}}$ (also called \textit{genon}) \cite{barkeshli2013genon}. This defect is terminated at the gapped boundary corresponding to the fold which condenses anyon pairs $e^{(1)}e^{(2)}$ or $m^{(1)}m^{(2)}$ of the bilayer $\Z_2$ toric code (as illustrated in Fig.~\ref{fig:folding_bilayer}(b)).
Note that the original long-range coupling in the cross-cap indicated by black lines in Fig.~\ref{fig:folding_bilayer}(a) is now turned into local inter-layer coupling in the bilayer twist indicated by pink dashed lines in Fig.~\ref{fig:folding_bilayer}(b).  The bilayer twist defect corresponds to the following automorphism for anyons going across it:
\begin{equation}
\rho_{S_{12}}: e^{(1)} \leftrightarrow e^{(2)}; \quad m^{(1)} \leftrightarrow m^{(2)}.    
\end{equation}

\subsubsection{Moving the layer-exchanging defect away from the fold}\label{subsec:moving_defect}

While the layer-exchanging defect is terminated at the gapped boundary in our folded code, one can think of moving the bilayer twist defects away from the gapped boundary corresponding to the fold, as shown in Fig.~\ref{fig:folding_bilayer}(c). The layer-exchanging defect then plays a role of a wormhole connecting the layers associated with the orientation-reversal.
The topology of the system is now turned into a sphere with two cross-caps instead of a single cross-cap, so the layer-exchanging defect can store two logical qubits.   In this double cross-cap geometry, the logical operators $\overline{X}_1$ and $\overline{Z}_1$ associated with the first qubit remains the same as those in the non-orientable toric code. Meanwhile, the logical operators $\overline{X}_2$ and $\overline{Z}_2$ associated with the second qubit go through the branch-cut of the bilayer twist and terminate at the $(e^{(1)}e^{(2)}, m^{(1)}m^{(2)})$ gapped boundary along the fold, as illustrated in Fig.~\ref{fig:folding_bilayer}(c).   If we apply the same emergent symmetry operator $U$ to this double cross-cap setup as in the case of a single cross-cap, we can get the corresponding logical gate as  $U=\overline{H}_1\overline{Z}_1 \otimes \overline{H}_2\overline{Z}_2$ on the logical qubits relevant to the layer-exchange defect.  

If we want to realize a logical Hadamard gate on the single logical qubit in this setup,  we can choose the encoding as $\ket{\overline{\psi}}_1 \otimes \ket{\overline{0}}_2$, where we have set the second logical qubit as an ancilla in the logical $0$ state.  Then when we trace out the second logical qubit, the symmetry operator $U$ acts as $\overline{H}\overline{Z}$ only on the single logical qubit. 

\subsubsection{Non-orientable surface code and its folding}

Though we originally defined the non-orientable toric code on a torus with a single cross-cap, we can also define the code with an open boundary with a cross-cap, i.e., a disk with a cross-cap with a choice of gapped boundary condition. We call the code with this setup as a ``non-orientable surface code''. This will ease the implementation of the logical Hadamard gate on the folded geometry, since the glued edges shown in Fig.~\ref{fig:folding_bilayer} cannot stay invariant when applying the stabilizer-pumping circuit and the solution is to simply replace them with open boundaries.    One potential choice of the gapped boundary of the folded non-orientable surface code with the boundary is to condense $(e^{(1)}e^{(2)}, m^{(1)}m^{(2)})$ on the boundary of the folded disk (see Fig.~\ref{fig:folding_bilayer}(b)). As shown in the lower panel of Fig.~\ref{fig:folding_bilayer}(b), the combination of the bilayer twist defects and a single type of gapped boundary $(e^{(1)}e^{(2)}, m^{(1)}m^{(2)})$ on the edge of a folded disk encodes a single logical qubit. By unfolding the picture, this choice of the gapped boundary would make up the geometry of the $\mathbb{RP}^2$ manifold, which is a sphere with a single cross-cap.
However, we note that it is not straightforward to implement this gapped boundary $(e^{(1)}e^{(2)}, m^{(1)}m^{(2)})$ on our non-orientable surface code in a way compatible with the action of the logical gates, and such gluing leads to additional resource cost in the hardware implementation.

In order to simplify the setup, we instead consider choosing the other specific gapped boundary of the disk, e.g., the $e$-condensed boundary which condense the $e$-anyons.   The corresponding unfolded code is illustrated in the left side of Fig.~\ref{fig:boundarytrans} where the entire boundary is chosen to be the $e$-boundary and a cross-cap is at the center.  After applying the emergent symmetry $U$, the entire boundary is switched to the $m$-condensed boundary, as shown in the right side of Fig.~\ref{fig:boundarytrans}. Folding the non-orientable surface code with the $e$-condensed boundary is described in Fig.~\ref{fig:folded_lattice}(b), which ends up with $(e^{(1)}, e^{(2)})$-condensed boundary of the bilayer $\Z_2$ toric code.  After applying the emergent symmetry operator $U$, this $(e^{(1)}, e^{(2)})$-condensed boundary is switched to  $(m^{(1)}, m^{(2)})$-condensed boundary.

While the boundary stabilizers shown in Fig.~\ref{fig:boundarytrans} is not a CSS code due to the presence of Pauli $Y$, one can bring it into a CSS code by applying $\exp(\frac{i\pi}{4}Z)$ or $\exp(-\frac{i\pi}{4}X)$ on boundary vertices where the Pauli $Y$ appears. After transforming it into a CSS code, one can obtain the logical CNOT gate acting on copies of these codes as shown in later discussions.

\subsubsection{Logical Hadamard gate in the presence of gapped boundary: formulating the non-orientable surface code as a subsystem code}

Let us consider the realization of the logical Hadamard gate of the non-orientable surface code in the presence of the gapped boundary. The issue here is that the symmetry $U$ switches the boundary type between $e$-condensed and $m$-condensed one, and hence does not define an automorphism of the original code. 

Meanwhile, one can regard $U$ as an automorphism of the code by re-formulating the non-orientable surface code as a subsystem code. To do this, let us express the stabilizer group of the non-orientable surface code with $e$-condensed boundary as $\mathcal{S}_e = \mathcal{S}_{\mathrm{bulk}} \times \mathcal{S}_{\mathrm{bdry}}^e$, where the stabilizers are separated into bulk and boundary stabilizers. Similarly, the stabilizer group of the non-orientable surface code with the $m$-condensed boundary is expressed as $\mathcal{S}_m = \mathcal{S}_{\mathrm{bulk}} \times \mathcal{S}_{\mathrm{bdry}}^m$. $U$ acts by an automorphism of the bulk stabilizer group $\mathcal{S}_{\mathrm{bulk}}$, while acting on the boundary stabilizer group as an isomorphism between $\mathcal{S}_{\mathrm{bdry}}^e$ and $\mathcal{S}_{\mathrm{bdry}}^m$.

Then, we construct a subsystem code by first setting the stabilizer group as $\mathcal{S}_{\mathrm{bulk}}$. In the subsystem code, some of the logical qubits in the stabilizer code $\mathcal{S}_{\mathrm{bulk}}$ does not store logical information, and regarded as ``gauge'' qubits that correspond to redundancy in encoding the quantum information. This gauge degree of freedom is specified by a gauge group $\mathcal{G}$, which is some subgroup of the normalizer $N(\mathcal{S}_{\mathrm{bulk}})$ of the stabilizer group $\mathcal{S}_{\mathrm{bulk}}$. In our case, we choose the gauge group to be $\mathcal{G} = (\mathcal{S}_{\mathrm{bdry}}^e\times\mathcal{S}_{\mathrm{bdry}}^m)/\tilde{\mathcal{S}}$, where $\tilde{\mathcal{S}}$ is a subgroup of $\mathcal{S}_{\mathrm{bulk}}$ that acts within the Hilbert space of boundary physical qubits. Concretely, $\tilde{\mathcal{S}}$ is generated by a pair of stabilizers given by the product of all $Z$ (resp.~$X$) stabilizers of $\mathcal{S}_{\mathrm{bulk}}$. 
The logical group of the subsystem code is then given by $N(\mathcal{S}_{\mathrm{bulk}})/\mathcal{G}$. This subsystem code stores a single logical qubit, whose logical Pauli operator is obtained by the $e,m$ string operator passing through the cross-cap.

One can then see that $U$ gives an automorphism of the subsystem code, since both $\mathcal{S}_{\mathrm{bulk}}$ and $\mathcal{G}$ are invariant under the action of $U$. This establishes the construction of the logical Hadamard gate in the subsystem code with a boundary.

In the above formalism in terms of the subsystem code, the choice of $e$ or $m$ condensed boundary condition for the non-orientable surface code corresponds to a different way of ``gauge fixing'' in the subsystem code, i.e., fixing the states of the gauge qubits. 
Practically, one needs a specific gauge fixing of the code by choosing one gapped boundary condition. After applying the operator $U$ on the code with the gauge fixing at the $e$-condensed boundary, $U$ changes the way of gauge fixing to the $m$-condensed boundary.
We then additionally need to perform $O(d)$ rounds of stabilizer measurements at the boundary to get back to the original gauge fixing.

\begin{table*}
\centering
\resizebox{1\columnwidth}{!}{%
	\begin{tabular}{|c|c|c|c|c|c|c|}
	\hline
	 	symmetry  & 
        setup &
        defect &
        locality     &  
        anyon permutation  & 
   logical gate   
	 	 \\
	 	\hline 
   \hline
spatial reflection&
unfolded  &
cross-cap / layer-exchange defect &
non-local&
no permutation &
N.A.
\\
\hline
layer exchange&
folded bilayer &
bilayer twist (genon) &
local and onsite&
$e^{(1)} \leftrightarrow e^{(2)}$, $m^{(1)} \leftrightarrow m^{(2)}$ &
N.A.
\\
\hline
$e$-$m$ exchange  &
unfolded &
anyon permutation &
local constant-depth circuit &
$e \leftrightarrow m$ &
$\overline{HZ}$
\\
\hline
transversal CZ  &
folded bilayer &
anyon permutation &
local and onsite&
$m^{(1)} \rightarrow m^{(1)}e^{(2)}, m^{(2)} \rightarrow m^{(2)}e^{(1)}$ &
$\overline{S}$
\\
\hline
transversal CNOT  &
copy of unfolded codes &
anyon permutation&
local and onsite&
$e^{1} \rightarrow e^{(1)}e^{(2)}$, $m^{(2)} \rightarrow m^{(1)}m^{(2)}$  &
$\overline{\text{CNOT}}$
\\
\hline
	 \end{tabular}
	 }
	 \caption{Summary of emergent symmetries, their corresponding defects and other properties in the folded or unfolded non-orientable code.}
	 \label{symmetry_table}
\end{table*}

\subsection{Generating the logical Clifford gate set on non-orientable toric code or surface code}
\label{subsec:clifford}
In addition to the logical Hadamard gate constructed above, one can obtain the full Clifford gate set on the copies of the non-orientable toric code or the subsystem code on a non-orientable open surface. All the Clifford gates are generated by a local constant-depth unitary circuit on the folded code (see Fig.~\ref{fig:folded_lattice} (b) for the folded code with $e$-condensed boundary). Below, we describe the realization of each generator of the Clifford group. 
The logical gates and emergent symmetries of the code including all Clifford gates, as well as their corresponding symmetry defects (both in the static setup and those related to the logical gates) are summarized in Table~\ref{symmetry_table}.

\subsubsection{Logical CNOT gate}
After one transforms the non-orientable toric code into a CSS code by the local unitary $\exp(\frac{i\pi}{4}Z_{0})\exp(-\frac{i\pi}{4}X_{2m})$ at the cross-cap, a transversal CNOT between two copies of the same type of codes leads to a logical CNOT gate:
\begin{equation}
\overline{\text{CNOT}} := \Motimes_{v\in \text{vertices}} \text{CNOT}_{1,2; v}.
\end{equation}

From the the field-theoretical perspective, the logical CNOT gate corresponds to the $\Z_2$ symmetry of the 
$\ZZ_2 \times \ZZ_2$ gauge theory that generates the outer automorphism of the gauge group, where $\ZZ_2 \times \ZZ_2$ represents two copies of the $\Z_2$ gauge theory. The symmetry action $\rho_{\text{CNOT}}$ on the anyons are given by
\begin{align}
\nonumber \rho_{\text{CNOT}}: & m^{(1)} \rightarrow  m^{(1)} m^{(2)}; m^{(2)} \rightarrow  m^{(2)}; \\
& e^{(2)} \rightarrow   e^{(2)} e^{(1)};  e^{(1)} \rightarrow  e^{(1)},
\end{align}
where $m^{(j)}$ ($e^{(j)}$) stands for the $m$
 ($e$) anyon in the $j^\text{th}$ copy.   Accordingly, the logical gate induces the following action on the Pauli string operators:
\begin{align}
\overline{\text{CNOT}} : & \overline{X}^{(1)} \rightarrow \overline{X}^{(1)}  \overline{X}^{(2)};   \overline{X}^{(2)}  \rightarrow \overline{X}^{(2)}; \\
& \overline{Z}^{(2)}  \rightarrow  \overline{Z}^{(2)}\overline{Z}^{(1)};  \overline{Z}^{(1)} \rightarrow  \overline{Z}^{(1)}.
\end{align}
The above transversal CNOT gate works as a logical gate even in the subsystem code on the surface, after we transform the code by local unitaries $\exp(\frac{i\pi}{4}Z)$ or $\exp(-\frac{i\pi}{4}X)$ on boundary vertices to bring it into a subsystem CSS code.

\subsubsection{Logical $S$ gate}
Here, we discuss the realization of the logical $S$ gate acting on the code space. Recall that the $S$ gate acts on the logical Pauli operators as $\overline X\to \overline Y, \overline Z\to \overline Z$. This action would correspond to the action on anyons given by $m\to\psi, e\to e$, which cannot be implemented by the local constant-depth circuit since it must preserve the spins of the anyons.
Below, we describe a realization of the logical $S$ gate on our code, which involves a non-local CZ gate acting on the physical qubits, but becomes local after folding the code in a certain fashion.

\begin{figure}
    \centering
    \includegraphics[width=1\textwidth]{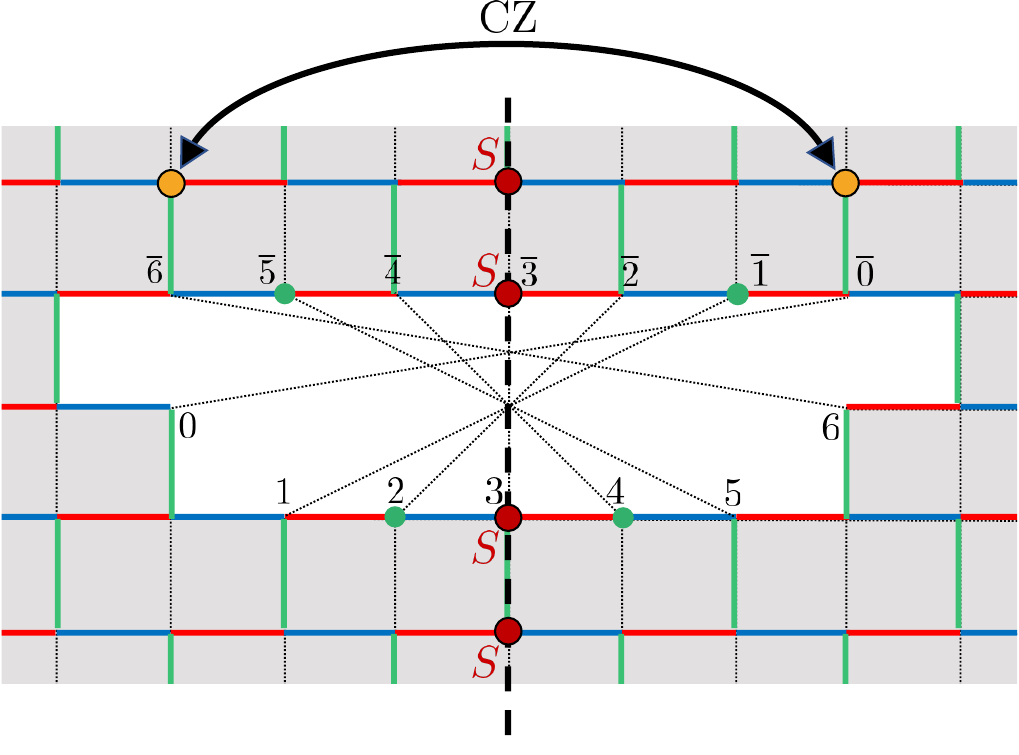}
\caption{Logical $S$ gate on the code with a cross-cap. It consists of the CZ gate for each pair of qubits related by the reflection along the central line, and the $S$ gate for each qubit on the central line.}
\label{fig:Sgatecrosscap}
\end{figure}

In order to implement the logical $S$ gate, we again work on the expression transformed into the CSS code by the local unitary $\exp(\frac{i\pi}{4}Z_{0})\exp(-\frac{i\pi}{4}X_{2m})$. We notice that the code has a reflection symmetry along the $y$ axis. We fold the code along the reflection line, and then apply transversal CZ gates on the aligned qubits and $S$ gates on the qubits on the fold, as illustrated in Fig.~\ref{fig:Sgatecrosscap}. 

In the folded system, the transversal CZ gate acts on the bulk by permuting the anyons of the layered system as
\begin{align}
\rho_{\text{CZ}}: m^{(1)} \rightarrow  m^{(1)} e^{(2)};   m^{(2)} \rightarrow  m^{(2)} e^{(1)},
\end{align}
while leaving $e^{(1)}, e^{(2)}$ invariant.
Also, the boundary of the folded system condenses $m^{(1)} m^{(2)}$ and $e^{(1)} e^{(2)}$, on which $\rho_{\text{CZ}}$ can be regarded as acting on anyons as $m\to m\times e, e\to e$; one can suppress the label of layers since the anyons on the different layers are identified on the boundary. Such an action is realized by the transversal $S$ gate on the boundary.

\begin{figure}
    \centering
    \includegraphics[width=1\textwidth]{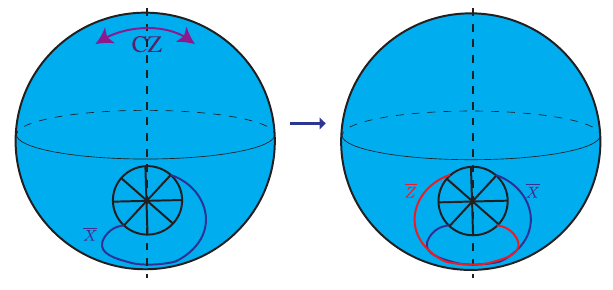}
\caption{The transversal CZ gate of the folded system acts on the Pauli $\overline{X}$ string operator by adding the $\overline{Z}$ string operator in the reflected position, realizing the action of the Pauli logical gate $\overline{S}:\overline{X} \rightarrow \overline{Y}$. }
\label{fig:foldRP2}
\end{figure}

The logical Pauli operators are $X$ or $Z$ string operators passing through the cross-cap exactly once. 
One can see that the logical gate indeed acts on these logical Pauli operators as
\begin{align}
\nonumber (\Motimes_{v \in \text{bulk}} \text{CZ}_{v})(\Motimes_{v \in \text{bdry}} S_v) : \quad  \overline{X} \rightarrow \overline{Y},
\end{align}
while keeping $\overline{Z}$ invariant. The action of the logical gate on the Pauli $\overline{X}$ is schematically shown in Fig.~\ref{fig:foldRP2}.

We note that the above gate works as a logical $S$ gate even in the case of the subsystem code on the surface, since the gate leaves the stabilizer and the gauge group invariant.~\footnote{Technically, we again need to apply the $\exp(-\frac{i\pi}{4}X)$ on boundary vertices involving the Pauli $Y$, to make the boundary stabilizers commute with the transversal CZ.}

\subsection{$\Z_2$ toric code on a Klein bottle with rotation symmetry and its logical gate}
\label{subsec:kleinbottle}

Here, we describe an alternative setup for the $\Z_2$ toric code on a Klein bottle.
See Fig.~\ref{fig:circular_code} for the illustration of the model.
This geometry is equivalent to a sphere with two cross-caps.  The tessellation of the north and south hemispheres are shown in Fig.~\ref{fig:circular_code}(a), where the white and black plaquettes again correspond to the $X$ and $Z$ stabilizers $S^X, S^Z$ respectively.  The center of both hemispheres contain the realization of cross-caps consisting of 4-body $X$ and $Z$ stabilizers connecting a pair of neighboring vertices and their inverted counterparts with respect to the central point.  The two hemispheres are glued along the equator by identifying the vertices with the same labels, from I to VIII.  It contains two logical qubits, with one pair of logical $X$ and $Z$ string operators travelling through each cross-cap. The logical Pauli operators stored by the north (resp.~south) cross-cap is denoted as $\{\overline{X}_1, \overline{Z}_1\}$ (resp.~$\{\overline{X}_2, \overline{Z}_2\}$).

Unlike the non-orientable toric code or surface code discussed in the previous subsections, the current geometry has the lattice rotation symmetry, which gives rise to an alternative realization of the logical Hadamard gate corresponding to exact microscopic symmetry.   This is in contrast to the logical Hadamard gate in the non-orientable toric code which is only an emergent symmetry.  However, due to the geometric constraint that the cross-cap in this case does not have a thin slit shape, there is no way to fold the Klein bottle into a bi-layer code as opposed to the case of the non-orientable toric code or surface code.    This illustrates the fact that generic emergent symmetries are more flexible than exact microscopic symmetries.  
Below, we describe the logical gates realized in the code on a Klein bottle.

\subsubsection{Logical Hadamard gate}\label{subsec:klein-bottle_hadamard}
\begin{figure*}
    \centering
    \includegraphics[width=0.9\textwidth]{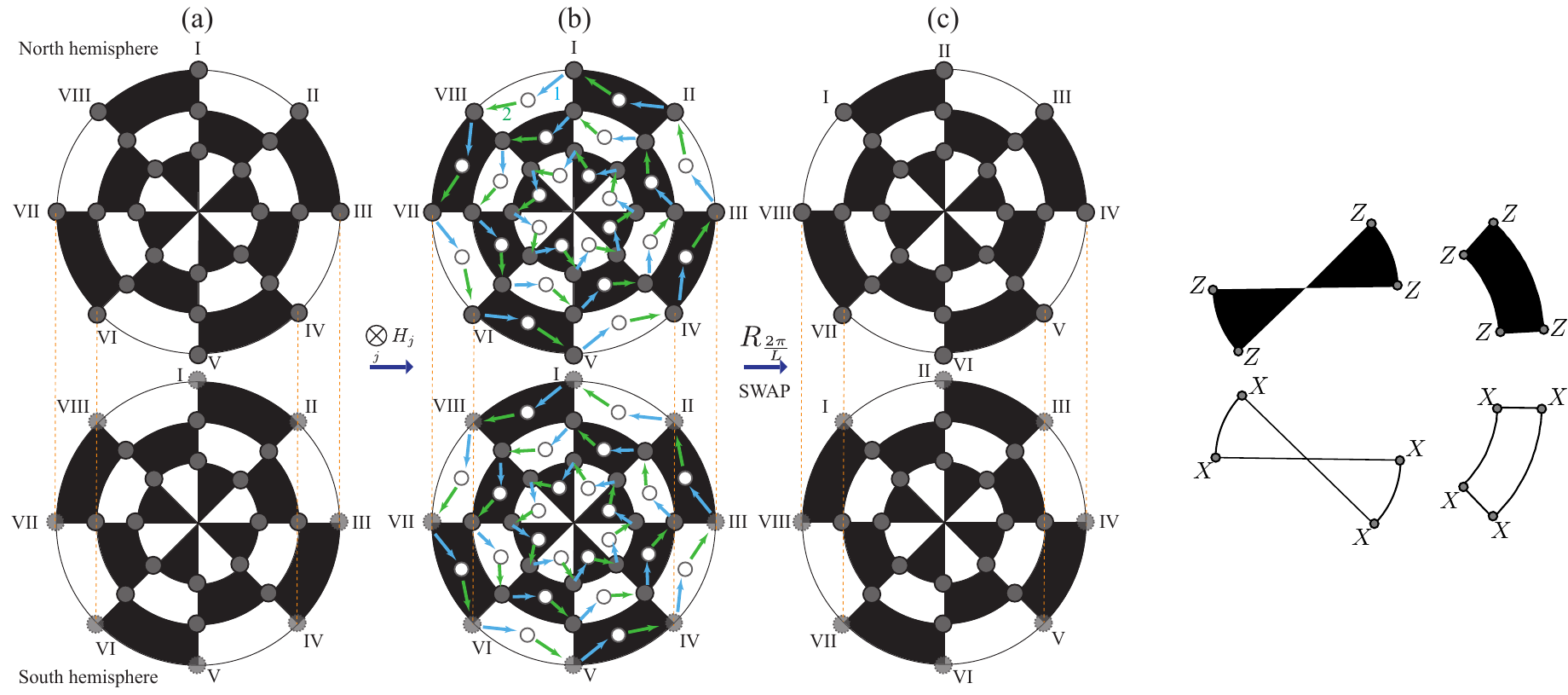}
\caption{The logical Hadamard gate of the $\Z_2$ toric code on a Klein bottle. (a): the Klein bottle can be regarded as a sphere with two cross-caps, and we locate the cross-caps at the north and south pole of the sphere. By applying the transversal Hadamard gate, the code is transformed to the one with the exchanged stabilizers $S^X\leftrightarrow S^Z$. (b): We then apply the rotation symmetry $R_{\frac{2\pi}{L}}$. The rotation operator can be expressed as the constant-depth local circuit by introducing the auxiliary qubits at the white dots, and then applying the local SWAP operators at the blue arrows firstly, and then at the green arrows. (c): After the combined transformation of the transversal Hadamard with the rotation, one can obtain the original stabilizer model. The combined transformation realizes the logical Hadamard gate $\overline{H}_1\otimes \overline{H}_2$.}
\label{fig:circular_code}
\end{figure*}

Since our lattice on the Klein bottle has the lattice symmetry that interchanges $S^X$ and $S^Z$, the $e\leftrightarrow m$ exchanging symmetry can easily be implemented by the transversal Hadamard gate associated with the spatial rotation.
Concretely, we firstly apply the transversal Hadamard gate $\Motimes_v H_v$, which interchanges the $S^X$ and $S^Z$ (black and white plaquettes) as well as the $X$ and $Z$ logical string operators:  
\begin{equation}
    \Motimes_v H_v : S^X_p \rightarrow S^Z_p; \quad S^Z_{p'} \rightarrow S^X_{p'};  \quad \overline{X}_{1.2} \leftrightarrow \overline{Z}_{1,2},
\end{equation}
 as illustrated in Fig.~\ref{fig:circular_code}. Since this operator alone interchanges the stabilizers $S^X\leftrightarrow S^Z$ and does not give a symmetry of the model, we need an additional operator to map the stabilizers back to original ones while inducing a trivial map on the logical operators. 
 As we can see, the Klein-bottle geometry in Fig.~\ref{fig:circular_code} has an $L$-fold rotation symmetry $R_{\frac{2\pi}{L}}$, where $L$ represents the number of qubits on each circle.  Therefore, the transversal Hadamard gate combined with the unit lattice rotation is a symmetry of the code space and also preserves the stabilizer group, while interchange the logical $X$ and logical $Z$ strings. This logical gate acts on the code space by the logical Hadamard gate on two qubits $\overline{H}_1\otimes \overline{H}_2$.  As has been discussed in Sec.~\ref{subsec:moving_defect}, we can set the second logical qubit as an ancilla in the state ${\overline{\ket{0}}}_2$ and trace it out. This will lead to the implementation of a single logical Hadamard gate $\overline{H}_1$

The rotation symmetry $R_{\frac{2\pi}{L}}$ can be implemented as the constant-depth local circuit by introducing auxiliary physical qubits, represented by white dots in Fig.~\ref{fig:circular_code} (b). The operator $R_{\frac{2\pi}{L}}$ can then be expressed as the product of local SWAP operators exchanging the black and white physical qubits as shown in Fig.~\ref{fig:circular_code} (b). In Appendix~\ref{sec:kleinmeasurement}, we demonstrate that the above Hadamard gate can also be implemented by a sequence of local Pauli measurements, reminiscent of the Floquet code.

\subsubsection{Logical CNOT and $S$ gate}
Since the above code on a Klein bottle has the CSS property, the code also admits implementation of the complete Clifford gate set, using the same method as the case with a single cross-cap in Sec.~\ref{subsec:clifford}.

If we prepare two copies of the code, a transversal CNOT between two copies of the same type of codes leads to a logical CNOT gate for each pair of qubits on a cross-cap:
\begin{equation}
\overline{\text{CNOT}}_1 \otimes \overline{\text{CNOT}}_2:= \Motimes_{v\in \text{vertices}} \text{CNOT}_{1,2; v}.
\end{equation}

In order to implement the logical $S$ gate, we again notice that the geometry has the reflection symmetry along the line connecting the vertex I and V in Fig.~\ref{fig:circular_code}, and then folding the code along the reflection line. After folding the code, applying the CZ gate in the bulk and $S$ gate on the boundary gives the logical $S$ gate,
\begin{align}
\overline{S}_1\otimes \overline{S}_2 = (\Motimes_{v \in \text{bulk}} \text{CZ}_{v})(\Motimes_{v \in \text{bdry}} S_v),
\end{align}
which is exactly the same construction as Fig.~\ref{fig:Sgatecrosscap}. 


\subsection{Error propagation for the logical gates}
\label{subsec:errorpropagation}

\begin{figure}
    \centering
    \includegraphics[width=1\textwidth]{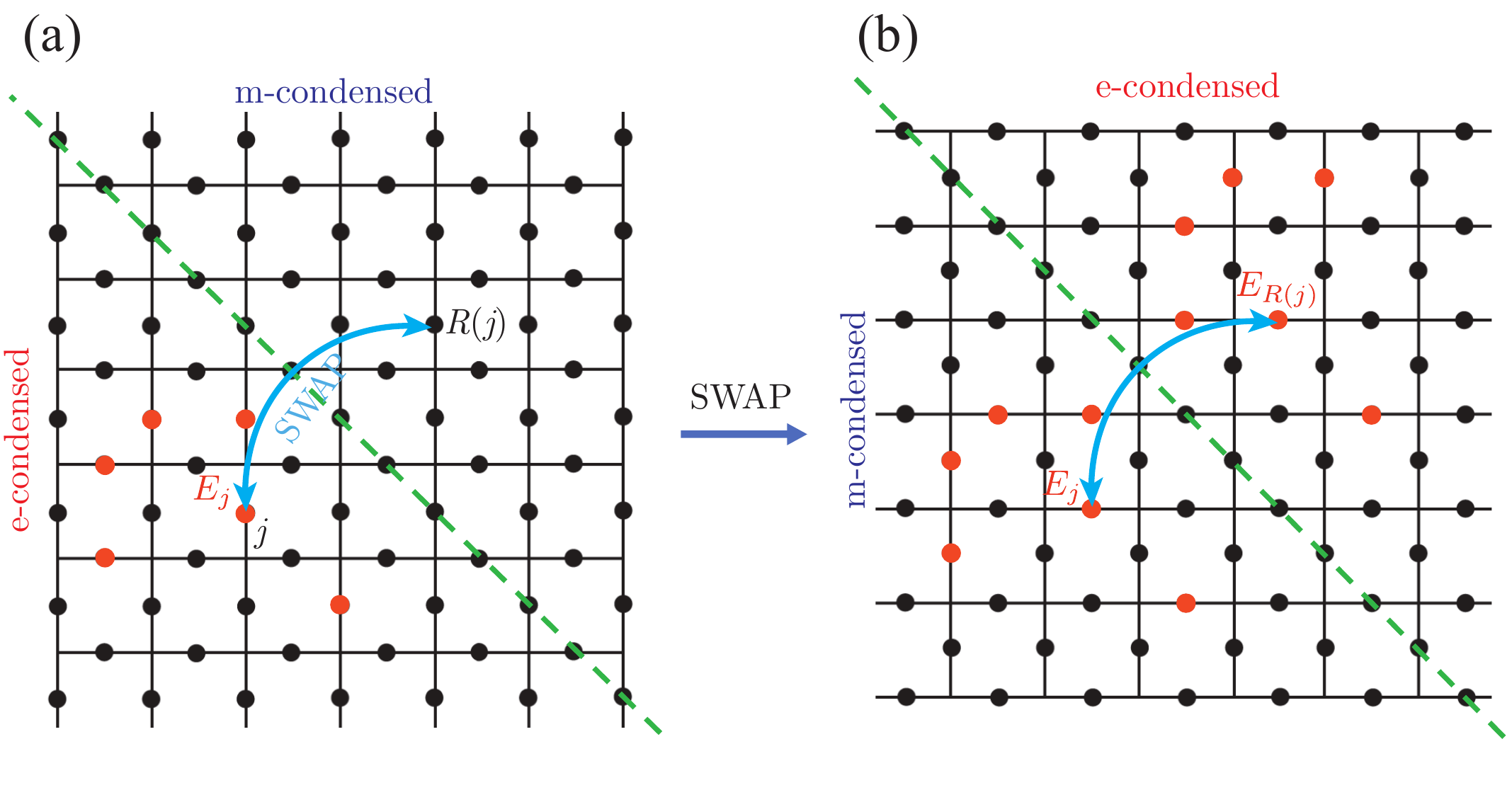}
\caption{Error propagation for the folded-transversal gate on the surface code.}
\label{fig:error_propagation}
\end{figure}

Now we compare the logical gates implemented via constant-depth circuits in the non-orientable toric code and the Klein-bottle code with the fold-transversal gates in the surface code discussed in Sec.~\ref{subsec:surface}.  

We note that the logical $H$ and $S$ gates corresponding to the fold-transversal gates in the surface code can propagate errors by a factor of two in general and can also induce long-range correlated errors.  We show this generic problem using the example of logical $H$ as illustrated in Fig.~\ref{fig:error_propagation}. It has been shown in Eq.~\eqref{eq:Hadamard_surface} that the logical Hadamard gate   $\overline{H} = \text{SWAP}\big(j \leftrightarrow R(j)\big) \Motimes_j H_j$  is composed by transversal Hadamard gate and pairwise long-range SWAPs with respect to the diagonal folding axis.  The transversal Hadamard gate $\Motimes_j H_j$ does not propagate any existing physical errors on the code, i.e., for any pre-existing error operator $E$, we have  $\Motimes_j H_j: E \rightarrow E'$
with $\textbf{supp} (E') = \textbf{supp}(E)$. On the other hand, if the pairwise SWAP gate is itself faulty with some small error rate $\epsilon$, it can map the error as 
\begin{align}\label{eq:error_folded}
\nonumber  &\text{SWAP}\big(j \leftrightarrow R(j)\big) : E' \rightarrow E'',  \\
  & \text{with} \ |\textbf{supp}(E'')| \le  2|\textbf{supp}(E')|,
\end{align}
which says that in the worst-case scenario the support size of the pre-existing error will be increased by a factor of two, as illustrated in  Fig.~\ref{fig:error_propagation}(a,b).
The errors are also propagated by a factor of two in the worst-case scenario for the logical-$S$ gate in the surface code via the fold-transversal CZ gate.    In both cases, the effective fault-tolerant code distance of the logical gate protocol can be reduced to approximately half of the original distance, i.e., $d_\text{eff} \approx \frac{1}{2}d$.   

Moreover, the introduction of the geometrically non-local coupling for the implementation of the long-range SWAP gates leads to long-range correlated errors.  Either considered as noise present for static non-local couplers or during the execution of the long-range SWAP gates, the noise of the qubits coupled together become long-range correlated.   One can hence characterize such noise via a noise Hamiltonian $H_{j,j'}$ where $j$ and $j'\equiv R(j)$ label the location of the two coupled qubits which are reflection symmetric with each other. In particular, Ref.~\cite{Aharonov:2006correlated} considers the situation where noise decays algebraically as
\begin{align}
||H_{j,j'}|| < \frac{\delta}{|j-j'|^z},  
\end{align}
where $|j-j'|$ denote the distance between the two coupled qubits.  It was shown in  Ref.~\cite{Aharonov:2006correlated} that there only exists an error threshold provided that $z>D$, where $D$ is the dimension of the qubit lattice (in the surface code case $D=2$).  Namely, fault-tolerant quantum computation is only scalable if the noise strength decays fast enough.    Now in the case of the fold-transversal gate, the noise strength scales as $||H_{j,j'}|| \sim \delta$, i.e., is independent of the distance between the two coupled qubits.   According to the study in Ref.~\cite{Aharonov:2006correlated}, it is likely that there may not be a fault-tolerant error threshold for the fold-transversal gate.   More detailed numerical analysis on the decoding will be required in future to understand the threshold behavior.

Now in terms of the logical Hadamard gate on the non-orientable toric codes, no long-range pairwise SWAP is needed as in the case of logical Hadamard gate in the surface code.    The logical gate corresponds to a local constant-depth circuit $U$, so the error propagation is constrained by a constant-size light cone \cite{Bravyi:2013dx}.   In particular, for any pre-existing error operator $E$ with support $\mathcal{R}=\textbf{supp}(E)$,  the error support after the action $\mathcal{R}'$$=$$\textbf{supp}(E')$$=$$\textbf{supp}(UEU^\dag)$ is contained within a $O(1)$-size neighborhood of $\mathcal{R}$. Also, to implement such constant-depth circuit, only gates and coupling between neighboring qubits on the lattice are needed and hence there is only short-range correlated noise.   This is in contrast to the long-range correlated noise of the fold-transversal gate.  Note that the only long-range connection needed in the non-orientable toric code is the static coupling along the cross cap. However, this is still considered to be local connectivity for the non-orientable toic code since it just connects qubits on the stabilizer support. Consequently, the presence of the short-range correlated noise in the non-orientable toric code will not affect the existence of a fault-tolerant error threshold for the implementation of the logical Hadamard gate.


We then compare the advantage/disadvantage of the logical Hadamard gate for the non-orientable toric codes versus the usual surface code in terms of detailed hardware implementation.  We first consider implementation with a single-layer (unfolded) system. In the case of the usual surface code, one needs $O
(d^2)$ long-range connection with $O(d)$-range to implement the long-range pairwise SWAPs and hence the logical Hadamard gate.     On the other hand, the non-orientable toric codes have logical Hadamard gates implemented by a constant-depth local circuit, so no long-range connection is needed for the the logical gate implementation.    The only long-range connection needed is static, which is needed for the engineering of the cross-cap in both cases and the number of long-range connections is $O(d)$.  It is expected that in most experimental architecture, the long-range coupling/gates have lower fidelity than short-range ones.  We can hence see the advantage of our scheme versus the usual surface code in terms of both the number of long-range connections ($O(d)$ versus $O(d^2)$) and the absence/presence of long-range correlated noise which potentially preserves/kills the error threshold.

Next, we compare the advantage/disadvantage in the folded bi-layer geometry in the context of the 3D pancake architecture.  As has been pointed out in Sec.~\ref{subsec:surface}, the long-range pairwise SWAP in the  logical Hadamard gate can be turned into geometrically local inter-layer SWAPs in the folded surface code, while this inter-layer coupling required to implement the SWAPs can also lead to long-range correlated noise on the surface code and absence of threshold.   For the non-orientable toric code on the Klein-bottle, we cannot fold it into a geometrically local code due to the shape of the cross-cap, as has been pointed out in Sec.~\ref{subsec:kleinbottle}.  On the other hand,  the non-orientable toric code with a slit-like cross-cap can be folded into a bi-layer local code such that the cross-cap can be implemented by local inter-layer connections, as pointed out in Sec.~\ref{subsec:folding}.   The logical Hadamard gate in this case is implemented by a constant-depth local circuit which does not need any inter-layer coupling when being away from the cross-cap. Therefore, the errors cannot be propagated into the other layer and no long-range correlated noise is present.  In this type of multi-layer architecture, it is expected that the intra-layer coupler and gates have higher fidelity than the inter-layer ones.   Therefore, from the hardware perspective, the fold-transversal logical Hadamard gate is expected to have the lower fidelity than the constant-depth logical Hadmard gate in the non-orientable gate since the inter-layer SWAP in the previous case is expected to have lower fidelity.    From the more fundamental perspective, the constant-depth logical Hadamard gate still has a fault-tolerant threshold due to the absence of long-range correlated noise, in contrast to the likely missing threshold in the case of the fold-transversal gate.

Now we also comment on some subtlety regarding the error propagation in both cases.   As has been discussed above, the error support is increased by a factor of two in the worst-case scenario for the fold-transversal gate, and hence leads to an effective distance reduced by half.  The situation of the constant-depth circuit for the non-orientable toric codes is more subtle.   For an error configuration with only a large connected error chain, the error support is only increased by $O(1)$ near the boundary of the connected error.  Therefore, for an error pattern consisting of a connected error chain with around $\sim d/2$ weight, the error propagation is negligible for the constant-depth circuit and remains correctable.  On the other hand,  the error weight may increase to $\sim d$ for the fold-transversal gate which now becomes uncorrectable.   Nevertheless, for an error pattern consisting of many sparse errors, the constant-depth circuit does not necessarily have an advantage over the fold-transversal gate.   This is because each small error cluster will increase by $O(1)$ near its boundary which will also lead to a constant factor increase of the total error weight.   We hence conclude that the constant-depth circuit has advantage in terms of the reduction of lower-fidelity long-range or inter-layer coupling/gates, the absence of long-range correlated noise and hence preservation of the fault-tolerant threshold, as well as the significantly surpressed error propagation for large connected error chain.

\begin{figure}
    \centering
    \includegraphics[width=0.9\textwidth]{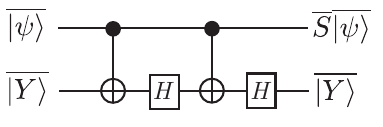}
\caption{The logical circuit implementing a logical $S$ gate with an input logical  state $\overline{\ket{Y}}$. }
\label{fig:logical_S_circuit}
\end{figure}

Finally, we note that, for the non-orientable toric codes, the logical CNOT gate is strictly a transversal gate and hence does not propagate errors. On the other hand, the logical $S$ gate in both cases is also a fold-transversal gate in both the unfolded and folded geometries, which  also has long-range correlated noise.   Therefore, it is expected that the logical $S$ gate has a lower fidelity than the logical $H$ and CNOT gates. In this case, we can use the less protected logical $S$ gate to inject a relatively more noisy $\overline{\ket{Y}}$ state.   Using a logical circuit composed of a higher-fidelity logical Hadamard and logical CNOT  as shown in Fig.~\ref{fig:logical_S_circuit}, one can implement the logical $S$ gate on the targeted logical qubit \cite{fowler2012}.  Now one can perform one or several rounds of state distillation, e.g., using the Steane code. The distillation circuit also uses the logical circuit in  Fig.~\ref{fig:logical_S_circuit} with the input $\overline{\ket{Y}}$ state from the last round of distillation (in the first round from the more noisy logical $S$ gate) \cite{fowler2012}.  With several rounds of such distillation process,  one can gradually improve the fidelity of the  $\overline{\ket{Y}}$ state and eventually implement a high-fidelity logical $S$ gate with the logical circuit in Fig.~\ref{fig:logical_S_circuit}.  In this way, we can achieve a high-fidelity logical Clifford gate set for the non-orientable toric codes.  One can further use the logical Clifford gate set to implement magic state distillation in order to achieve universal fault-tolerant quantum computation with the non-orientable code.

\section{Logical gate of honeycomb Floquet code}
\label{sec:floquet}
So far, we studied the logical gates of the stabilizer code, where the code space is implemented by the ground states of a gapped Hamiltonian.
Here we consider an alternative setup of implementing the code space by a sequence of measurements rather than a specific Hamiltonian, known under the name of the Floquet code~\cite{Hastings2021honeycomb}. 

Analogously to what we have done in Sec.~\ref{sec:RP2code} for the $\Z_2$ toric code, we define a honeycomb Floquet code on a surface with a single cross-cap, and show that the a period of the sequence for measurements implements the $\exp(-\frac{i\pi}{4}\overline{Y}) = \overline{H}\overline{Z}$ logical gate on the code space. 

\subsection{Review: honeycomb Floquet code without cross-cap}
\label{sec:honeycombreview}

\begin{figure}
    \centering
    \includegraphics[width=0.8\textwidth]{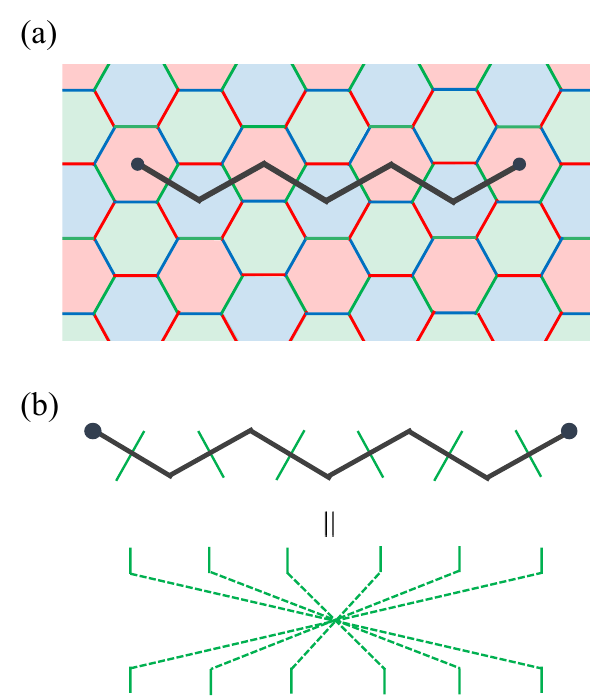}
\caption{(a): the hexagonal lattice with the plaquettes and edges colored by R, G, and B. The 2d space is a torus with the periodic boundary condition along the horizontal and vertical directions.
When we want to introduce a cross-cap to the 2d space, we locate it along the thick black line. (b): the cross-cap connects the edges cutting it to those at the reflected position, which makes up a non-orientable surface. Note that the action of the cross-cap preserves the coloring of plaquettes and edges, so the coloring makes sense in the presence of the cross-cap. Though the figure represents the cross-cap with length seven, one can easily generalize the cross-cap to have general odd length.}
\label{fig:coloredhexagon}
\end{figure}

Let us first review the honeycomb Floquet code on a 2d torus introduced in Ref.~\cite{Hastings2021honeycomb}.
The model is defined on a honeycomb lattice with single qubit on each vertex. We take periodic boundary condition of the honeycomb lattice, and color the edges and faces as shown in Fig.~\ref{fig:coloredhexagon}.
The honeycomb Floquet code is defined by a sequence of measurements instead of preparing a fixed Hamiltonian. We label each round of measurement by an integer $r\ge 0$, and measure the operators called ``checks'' on edges in the form of $XX$, $YY$, $ZZ$ in the following fashion with periodicity mod 3,
\begin{itemize}
    \item When $r=0$ mod 3, we measure the check $XX$ on R edges.
    \item When $r=1$ mod 3, we measure the check $YY$ on G edges.
    \item When $r=2$ mod 3, we measure the check $ZZ$ on B edges.
\end{itemize}
This completes the definition of the honeycomb Floquet code. Though there is no specific Hamiltonian that protects the code space, the sequence of measurements gives a certain stabilizer group protecting the logical qubit after each round $r$, which is called an instantaneous stabilizer group $\mathcal{S}_r$ labeled by the round $r$. 
Let us write the measurement outcome of a check at the round $r$ as $v_{ij}^r$ for each edge $\langle ij\rangle$.
Then, $\mathcal{S}_r$ for first several rounds are described as follows. 
\begin{itemize}
    \item After the round $r=0$, $\mathcal{S}_0$ is generated by $v_{ij}^0 X_iX_j$ on each R edge $\langle ij\rangle$. 
     \item After the round $r=1$, $\mathcal{S}_1$ is generated by $v_{ij}^1 Y_iY_j$ on each G edge $\langle ij\rangle$, together with the operator $w_p W_p$ on each B plaquette $p$, where $w_p\in\{\pm 1\}$ and the operator $W_p$ on a B plaquette $p$ is defined as
     \begin{align}
         w_p = \prod_{\langle ij\rangle \subset \partial p} v^r_{ij}, \quad W_p = \prod_{\text{G edge}\subset\partial p} Y_iY_j  \prod_{\text{R edge}\subset\partial p} X_iX_j 
         \end{align}
         where $r=0,1$ depending on the color of edges.
    \item After the round $r=2$, $\mathcal{S}_2$ is generated by $v_{ij}^2 Z_iZ_j$ on each B edge $\langle ij\rangle$, together with the operator $w_p W_p$ on each R, B plaquette $p$, where $w_p\in\{\pm 1\}$ and the operator $W_p$ on a R plaquette $p$ is defined as  
\begin{align}
         w_p = \prod_{\langle ij\rangle \subset \partial p} v^r_{ij}, \quad W_p = \prod_{\text{B edge}\subset\partial p} Z_iZ_j  \prod_{\text{G edge}\subset\partial p} Y_iY_j 
         \end{align}
         where $r=1,2$ depending on the color of edges.
         
         \item After the round $r=3$, $\mathcal{S}_3$ is generated by $v_{ij}^3 X_iX_j$ on each R edge $\langle ij\rangle$, together with the operator $w_p W_p$ on each R, G, B plaquette $p$, where $w_p\in\{\pm 1\}$ and the operator $W_p$ on a G plaquette $p$ is defined as   
\begin{align}
         w_p = \prod_{\langle ij\rangle \subset \partial p} v^r_{ij}, \quad W_p = \prod_{\text{R edge}\subset\partial p} X_iX_j  \prod_{\text{B edge}\subset\partial p} Z_iZ_j 
         \end{align}
         where $r=2,3$ depending on the color of edges.
         
\end{itemize}
After the third round of the measurement $r=3$, the instantaneous stabilizer group contains the operators $\{w_p W_p\}$ for all R, G, B plaquettes. These plaquette operators commute with any checks on edges, so they are permanent elements of the stabilizer group $\mathcal{S}_r$ for any $r\ge 3$. The group $\mathcal{S}_r$ for $r\ge 3$ is then given as follows.
\begin{itemize}
    \item When $r=0$ mod 3 and $r\ge 3$, $\mathcal{S}_r$ is generated by $\{w_p W_p\}$ for all plaquettes $p$, together with $v_{ij}^r X_i X_j$ on each R edge. On each G plaquette $p$, the measurement outcomes are subject to the constraint 
\begin{align}
    \prod_{\text{R edge}\subset\partial p} v_{ij}^r \prod_{\text{B edge}\subset\partial p} v_{ij}^{(r-1)} = w_p.
    \end{align}
    
    \item When $r=1$ mod 3 and $r\ge 3$, $\mathcal{S}_r$ is generated by $\{w_p W_p\}$ for all plaquettes $p$, together with $v_{ij}^r Y_i Y_j$ on each G edge.
On each B plaquette $p$, the measurement outcomes are subject to the constraint 
\begin{align}
    \prod_{\text{G edge}\subset\partial p} v_{ij}^r \prod_{\text{R edge}\subset\partial p} v_{ij}^{(r-1)} = w_p.
    \end{align}
    
    \item When $r=2$ mod 3 and $r\ge 3$, $\mathcal{S}_r$ is generated by $\{w_p W_p\}$ for all plaquettes $p$, together with $v_{ij}^r Z_i Z_j$ on each B edge. 
On each R plaquette $p$, the measurement outcomes are subject to the constraint 
\begin{align}
    \prod_{\text{B edge}\subset\partial p} v_{ij}^r \prod_{\text{G edge}\subset\partial p} v_{ij}^{(r-1)} = w_p.
    \end{align}
    
    \end{itemize}
For $r\ge 3$, the stabilizer group $\mathcal{S}_r$ stores two qubits, where the code space is regarded as the effective $\Z_2$ gauge theory ($\Z_2$ toric code). The code space of the stabilizer group $\mathcal{S}_r$ is denoted as $\mathcal{C}_r$. 

\subsection{Honeycomb Floquet code in the presence of a single cross-cap}

Here we consider the honeycomb Floquet code on a torus in the presence of a cross-cap. The cross-cap is implemented in the hexagonal lattice as shown in Fig.~\ref{fig:coloredhexagon}. 
Note that the cross-cap is chosen so that it preserves the coloring of hexagons and edges. The honeycomb Floquet code can then be described in the exactly same fashion as Section \ref{sec:honeycombreview} based on the colors of the edges.

The only extra thing we need to care about in the presence of the cross-cap is an exceptional plaquette with length 12 at the end of a cross-cap. After the round of the measurement at $r=2$, the plaquette stabilizer $w_p W_p$ for this exceptional plaquette is also introduced in the instantaneous stabilizer group $\mathcal{S}_2$, which is given by
\begin{align}
         w_p = \prod_{\langle ij\rangle \subset \partial p} v^r_{ij}, \quad W_p = \prod_{\text{B edge}\subset\partial p} Y_iY_j  \prod_{\text{G edge}\subset\partial p} X_iX_j 
         \end{align}
where $p$ is the plaquette with length 12, and $r=1,2$ depending on the color of edges. This plaquette stabilizer is an element of $\mathcal{S}_r$ for any $r\ge 2$, and the measurement outcomes for $r=2$ mod 3 follows the constraint at the plaquette
\begin{align}
    \prod_{\text{B edge}\subset\partial p} v_{ij}^r \prod_{\text{G edge}\subset\partial p} v_{ij}^{(r-1)} = w_p.
    \end{align}

\subsection{Logical gate implemented by dynamics}
Here we evaluate the logical gate implemented by the sequence of the three steps of measurements, starting at the $r$-th round with $r=0$ mod 3. 
The argument of this section applies for the honeycomb Floquet code with or without cross-cap.

We need to fix the basis of the code space $\mathcal{C}_{r}$ and $\mathcal{C}_{r+3}$ to talk about the logical gate acting on the code space $\mathcal{C}_{r}$, which will be considered later.
Once the basis states are specified, the logical gate for the sequence of three measurements at $r+1,r+2,r+3$ is given by the matrix element
\begin{align}
    \bra{\psi_{r+3}} \Pi_{r+2} \Pi_{r+1} \ket{\psi_{r}}
\end{align}
with some basis state $\ket{\psi_r}\in \mathcal{C}_r$, $\ket{\psi_{r+3}}\in \mathcal{C}_{r+3}$, where $\Pi_{r+1}, \Pi_{r+2}$ are projectors onto the measurement outcome given by
\begin{align}
\begin{split}
    \Pi_{r+1} &= \prod_{\langle ij\rangle\in\text{G edges}}\frac{1+ v^{(r+1)}_{ij}Y_iY_j}{2}, \\ \Pi_{r+2} &= \prod_{\langle ij\rangle\in\text{B edges}}\frac{1+ v^{(r+2)}_{ij}Z_iZ_j}{2}.
    \end{split}
\end{align}
By expanding the sum of the projectors, one can express the operator $\Pi_{r+2} \Pi_{r+1}$ as the sum of the operators in the form of
\begin{align}
\begin{split}
\Pi_{r+2} \Pi_{r+1}=\frac{1}{2^{2N}}\sum_{C_{\text{GB}}}&\left(\prod_{\text{B edges}\in C_{\text{GB}}}v^{(r+2)}_{ij}Z_iZ_j \right) \\
\times & \left(\prod_{\text{G edges}\in C_{\text{GB}}}v^{(r+1)}_{ij}Y_iY_j \right)
\label{eq:PiPisum}
\end{split}
    \end{align}
where $N$ is the number of unit cells, and
$C_{\text{GB}}$ is set of edges that contains only G, B edges. We sum over all possible choices of such a set $C_{\text{GB}}$. For a given set $C_{\text{GB}}$, it is convenient to label each R edge $e$ by an integer $0,1,2$, depending on how many vertices overlap between $\partial(C_{\text{GB}})$ and $\partial e$.\footnote{Here we regard $C_{\text{GB}}$ as a $\Z_2$-valued 1-chain which is sum over edges of $C_{\text{GB}}$. We sometimes use this abuse of notation.} Let us denote the set of all R edges with label $j$ as $C_{\text{R},j}$. 
Then, the operator in the summand of Eq.~\eqref{eq:PiPisum} anti-commutes with the $XX$ check on an R edge $e$, if and only if $e$ is contained in $C_{\text{R},1}$. So, the summand has a non-zero matrix element between $\mathcal{C}_r$ and $\mathcal{C}_{r+3}$ if and only if $C_{\text{R},1}$ coincides with the set of R edges with $v^r_e = - v^{(r+3)}_e$. Let us denote this set of R edges as $\hat C_{\text{R}}$, then within the code space $\mathcal{C}_r$ we have
\begin{align}
\begin{split}
\Pi_{r+2} \Pi_{r+1}=\frac{1}{2^{2N}}
\sum_{\substack{C_{\text{GB}} \\ C_{\text{R},1} =\hat C_{\text{R}}}}  & \left(\prod_{\text{B edges}\in C_{\text{GB}}}  v^{(r+2)}_{ij}Z_iZ_j \right) \\
\times\left(\prod_{\text{G edges}\in C_{\text{GB}}}v^{(r+1)}_{ij}Y_iY_j \right)\times &\left(\prod_{\text{R edges}\in C_{\text{R},2}}v^{r}_{ij}X_iX_j \right)
\end{split}
\end{align}
Here, the operator is supported on the set of edges $C_{\text{GB}} + C_{\text{R},2}$, which constitutes an open line operator with the end at edges of $\hat C_{\text{R}}$. This is physically regarded as an operator creating the fermionic particle at the location of $\hat C_{\text{R}}$. Let us simply denote $C'=C_{\text{GB}} + C_{\text{R},2}$, then we get
\begin{align}
\Pi_{r+2} \Pi_{r+1}=\frac{1}{2^{2N}}
\sum_{\substack{C' \\ C_{\text{R},1} =\hat C_{\text{R}}}} V(C'),
\label{eq:pi2pi1}
\end{align}
where $C'$ is an open curve that satisfies the property that $C_{\text{R},1} =\hat C_{\text{R}}$. 
$V(C')$ is the open line operator for the fermionic particle
\begin{align}
\begin{split}
V(C')=&\left(\prod_{\text{B edges}\in C'}v^{(r+2)}_{ij}Z_iZ_j \right) \\
\times & \left(\prod_{\text{G edges}\in C'}v^{(r+1)}_{ij}Y_iY_j \right) \left(\prod_{\text{R edges}\in C'}v^{r}_{ij}X_iX_j \right).
\end{split}
\end{align}
In Appendix~\ref{app:invarianceVC}, we show that the action of $V(C')$ within the code space is invariant under shifting $C'$ by boundary $C'\to C' + \partial F$, where $F$ is any set of plaquettes. That is, one can show that
\begin{align}
    \bra{\psi_{r+3}}V(C'+\partial F)\ket{\psi_r} = \bra{\psi_{r+3}}V(C')\ket{\psi_r},
    \label{eq:invariance_under_dFshift}
\end{align}
for any set of plaquettes $F$.

Then, let us rewrite the open curve $C'$ as $C'=C+C_0$, where $C_0$ is a fixed set of open curves satisfying $C_{\text{R},1} =\hat C_{\text{R}}$, and $C$ is a closed loop. Then, the sum over $C'$ can be rewritten as the sum over the closed loops $C$. Then we have
\begin{align}
\begin{split}
    \bra{\psi_{r+3}} \Pi_{r+2} \Pi_{r+1}\ket{\psi_r} &= \\
    \frac{2^{|F|-1}}{2^{2N}}  \sum_{[C]\in H_1(\Sigma,\Z_2)} & \bra{\psi_{r+3}}V(C + C_0)\ket{\psi_r},
    \end{split}
\end{align}
where $|F|$ is the number of plaquettes; $|F|= N$ for the torus without a cross-cap, while $|F|= N-1$ with a cross-cap.
$\Sigma$ is a 2d surface supporting a honeycomb Floquet code, i.e., a torus with or without a cross-cap.

The sum over closed loops $C$ is written as the sum over homology classes $[C]\in H_1(\Sigma,\Z_2)$, since the summand only depends on the homology class of $C$ due to Eq.~\eqref{eq:invariance_under_dFshift}.
For a fixed choice of $C_0$ given by a set of open curves, one can always find a representative $C$ for $[C]\in H_1(\Sigma,\Z_2)$ satisfying the property $V(C+C_0) = V(C)V(C_0)$.~\footnote{Here, we are choosing a representative $C$ of $[C]\in H_1(\Sigma,\Z_2)$ such that $\sharp\mathrm{int}(C,C_0)=0$, where $\sharp\mathrm{int}(C,C_0)$ the mod 2 intersection number.
For a general choice of representative $C$ of $[C]\in H_1(\Sigma,\Z_2)$ we have $V(C+C_0) = V(C)V(C_0)(-1)^{\sharp\mathrm{int}(C,C_0)}$ (see Appendix~\ref{app:quadratic}). One can then shift the representative $C$ of $[C]\in H_1(\Sigma,\Z_2)$ by boundary of some region to shift $\sharp\mathrm{int}(C,C_0)$ by some odd number, so that $C$ after the modification satisfies $V(C+C_0) = V(C)V(C_0)$.} Based on these representatives $C$, we have
\begin{align}
\begin{split}
    \bra{\psi_{r+3}} \Pi_{r+2} \Pi_{r+1}\ket{\psi_r} &= \\
    \frac{2^{|F|-1}}{2^{2N}} \sum_{[C]\in H_1(\Sigma,\Z_2)} & \bra{\psi_{r+3}}V(C) V(C_0)\ket{\psi_r}.
    \end{split}
\end{align}
Now we can compute the action of the logical gate. Suppose that the basis of  $\mathcal{C}_r, \mathcal{C}_{r+3}$ are eigenstates of the line operators $V(C)$ along the non-contractible closed loop $C$. For each basis state $\ket{\psi_r}$ of $\mathcal{C}_r$, we define the basis state of $\mathcal{C}_{r+3}$ with the same eigenvalues as $\ket{\psi_{r+3}} := V(C_0)\ket{\psi_r}$. Here, $V(C_0)\ket{\psi_r}$ is a state of the code space $\mathcal{C}_{r+3}$ because $V(C_0)$ gives an automorphism between instantaneous stabilizer groups $\mathcal{S}_r$ and $\mathcal{S}_{r+3}$, i.e., $V(C_0)\mathcal{S}_rV(C_0)^\dagger = \mathcal{S}_{r+3}$. 
Then, the logical gate is diagonal and given by
\begin{align}
    \frac{2^{|F|-1}}{2^{2N}} \sum_{[C]\in H_1(\Sigma,\Z_2)} \bra{\psi_{r+3}}V(C)\ket{\psi_{r+3}}.
\label{eq:logicalArf}
\end{align}
This can be computed by using the quadratic property (see Appendix \ref{app:quadratic} for a proof)
\begin{align}
    V(C+C') = (-1)^{\sharp\mathrm{int}(C,C')}  V(C)V(C'),
\end{align}
where $C,C'$ are closed loops and $\sharp\mathrm{int}(C,C')$ is the mod 2 intersection number. When $\ket{\psi_{r+3}}$ is an eigenstate of $V(C)$, $\langle V(C)\rangle := \bra{\psi_{r+3}}V(C)\ket{\psi_{r+3}}$ gives a quadratic form
\begin{align}
\langle V(C+C')\rangle = (-1)^{\sharp\mathrm{int}(C,C')}\langle V(C)\rangle\langle V(C')\rangle.
\end{align}
It is known that the following sum of this quadratic form becomes a phase
\begin{align}
    \frac{1}{\sqrt{|H_1(\Sigma,\Z_2)|}}\sum_{[C]\in H_1(\Sigma,\Z_2)} \langle V(C)\rangle,
\end{align}
which is called the Arf invariant of the quadratic form $\langle V(C)\rangle$ when $\Sigma$ is orientable, while it is called the Arf-Brown-Kervaire (ABK) invariant when $\Sigma$ is non-orientable~\cite{Brown1972}. This Arf or ABK invariant gives the diagonal matrix element of the logical gate Eq.~\eqref{eq:logicalArf}. While the Arf invariant for oriented surface takes value in $\{\pm 1\}$, the ABK invariant for a non-orientable surface $\Sigma$ is valued in $e^{\frac{2\pi i}{8}\nu}$ with $\nu\in\Z_8$. Thanks to this eighth root of unity, one can see that the non-orientable geometry enriches the dynamics of the honeycomb Floquet code.

Below, let us explicitly describe the logical gate implemented by a period of measurements for a torus with or without a cross-cap.

\begin{itemize}
    \item When the 2d surface $\Sigma$ is a torus without cross-cap, one can write 
    \begin{align}
        \langle V(C_x)\rangle=s_x, \quad \langle V(C_y)\rangle=s_y,
        \end{align}
with $C_x$, $C_y$ the non-contractible 1-cycles, and we obtain the final expression
\begin{align}
    \bra{\psi_{r+3}} \Pi_{r+2} \Pi_{r+1}\ket{\psi_r} = \frac{1}{2^N}\mathrm{Arf},
\end{align}
\begin{align}
    \mathrm{Arf}= \frac{1+s_x+s_y-s_xs_y}{2}.
\end{align}
Here, $\mathrm{Arf}\in\{\pm 1\}$ is the Arf invariant of the quadratic form $\bra{\psi_{r+3}} V(C)\ket{\psi_{r+3}}$.
If we instead work on a standard choice of basis where the fermionic line operators $V(C_x), V(C_y)$ of the $\Z_2$ toric code are identified as the $\overline{X}_1\overline{Z}_2$, $\overline{X}_2\overline{Z}_1$ gates,
the operator $\mathrm{Arf}$ is given by
\begin{align}
    \mathrm{Arf} = \frac{1+\overline{X}_1\overline{Z}_2+\overline{X}_2\overline{Z}_1 + \overline{X}_1\overline{Z}_1 \overline{X}_2\overline{Z}_2}{2}.
    \label{eq:Arfoperator}
\end{align}
One can check that Arf transforms the Pauli operators as
\begin{align}
    \overline{X}_1\leftrightarrow \overline{Z}_2, \quad \overline{X}_2\leftrightarrow \overline{Z}_1
\end{align}
that is, it implements the $e\leftrightarrow m$ exchange $\Z_2$ symmetry of the $\Z_2$ gauge theory.
 \item  When the 2d surface $\Sigma$ is a torus with a single cross-cap, one can write
 \begin{align}
    \langle V(C_x)\rangle=s_x,  \langle V(C_y)\rangle=s_y,  \langle V(C_w)\rangle=s_w \end{align}
where $C_w$ is the 1-cycle that crosses through a cross-cap once.
 Note that $s_w = \pm i$, since $C_w$ has self-intersection and $V(C_w)V(C_w) = -1$.
 Then we obtain the final expression
 \begin{align}
     \bra{\psi_{r+3}} \Pi_{r+2} \Pi_{r+1}\ket{\psi_r} = \frac{\sqrt{2}}{2^N}\cdot \mathrm{ABK}, 
     \end{align}
     \begin{align}
    \mathrm{ABK}= \frac{1+s_x+s_y-s_xs_y}{2}\cdot \frac{1+s_w}{\sqrt{2}}
    \label{eq:ABK}
\end{align}
Here, $\mathrm{ABK}$ is the Arf-Brown-Kervaire invariant of the quadratic form $\langle V(C)\rangle$ which is valued in $e^{\frac{2\pi i}{8}\nu}$ with $\nu\in\Z_8$.

If we instead work on the standard basis of $\Z_2$ toric code where the fermionic line operators $V(C_x), V(C_y)$ are identified as the $\overline{X}_1\overline{Z}_2$, $\overline{X}_2\overline{Z}_1$ and $V(C_w)$ is identified as $\overline{X}_3\overline{Z}_3 = -i\overline{Y}_3$, the operator ABK is given by the matrix representation
\begin{align}
    \mathrm{ABK} = \mathrm{Arf}_{1,2}
    \otimes \exp\left(-\frac{i\pi}{4}\overline{Y}_3\right),
    \label{eq:logicaldynamics_crosscap}
\end{align}
where Arf$_{1,2}$ is the Arf operator in Eq.~\eqref{eq:Arfoperator} for the first two qubits implemented on a torus, and $\exp\left(-\frac{i\pi}{4}\overline{Y}_3\right) = \overline{H}_3\overline{Z}_3$ is a Clifford gate for a third qubit encoded by a cross-cap. This gate generates the $\Z_4$ group up to phase, $(\overline{H}_3\overline{Z}_3)^2= -i \overline{Y}_3, (-i \overline{Y}_3)^2 = -1$. \footnote{Though the expression of the logical gate Eq.~\eqref{eq:logicaldynamics_crosscap} depends on a choice of basis of the code space, we remark that the conclusion that it generates the $\Z_4$ group is not affected by redefinition of the basis $\ket{\psi_{r+3}}= V(C_0)\ket{\psi_r}$ shifting $C_0$ by a closed loop $C_0\to C_0 + C'$ with $[C']\in H_1(\Sigma,\Z_2)$. This redefinition does not change the eigenvalues of $\ket{\psi_{r+3}}$, and amounts to shifting the expression of the logical gate Eq.~\eqref{eq:logicaldynamics_crosscap} by multiplying $V(C')$. In particular, if $C'= C_w$, it shifts $\exp\left(-\frac{i\pi}{4}\overline{Y}_3\right)\to -\exp\left(\frac{i\pi}{4}\overline{Y}_3\right)$, so it still generates the $\Z_4$ group.}

\end{itemize}
\

\subsection{Dynamics as a condensation operator of $\Z_2$ gauge theory}
\label{subsec:condensation}
Here we describe the dynamics of the honeycomb Floquet code in terms of effective $\Z_2$ gauge theory. In the above discussions, we have seen that a period of the dynamics implements a logical gate given by sum over insertions of fermionic line operators $V(C)$ proportional to
\begin{align}
    \frac{1}{\sqrt{|H_1(\Sigma,\Z_2)|}}\sum_{[C]\in H_1(\Sigma,\Z_2)}  V(C).
    \label{eq:psicondensation}
\end{align}
At the level of effective field theory, an operator supported on a surface $\Sigma$ obtained by summing over possible insertions of an Wilson line operator in $\Sigma$ is called a condensation operator~\cite{seifnashri2022condensation}. The operator in Eq.~\eqref{eq:psicondensation} gives a lattice description for the condensation operator of a fermion $\psi$ supported on a 2d space $\Sigma$. This condensation operator of (2+1)D $\Z_2$ gauge theory is known to generate 0-form invertible symmetry that exchanges anyons $e\leftrightarrow m$~\cite{seifnashri2022condensation}. 

The group structure generated by this condensation operator depends on whether the 2d surface $\Sigma$ is orientable or not.
To see this, we note that the 2d condensation operator for the fermion $\psi$ in $\Z_2$ gauge theory corresponds to a symmetry defect obtained by gauging $\Z_2^f$ symmetry of (1+1)D Kitaev's Majorana chain located at a subsystem of (2+1)D trivial atomic insulator~\cite{Barkeshli2023codim2}. That is, this condensation operator is described in (2+1)D spacetime by starting with a (2+1)D trivial fermionic insulator with $\Z_2^f$ symmetry with a location of Kitaev chain at a codimension-1 submanifold, and then gauging $\Z_2^f$ symmetry of the whole system. After gauging $\Z_2^f$ symmetry, the Kitaev chain is realized as a codimension-1 condensation defect of the $\Z_2$ gauge theory.

From this perspective, one can see that the fusion rule of the condensation operators reflects the stacking law of the Kitaev chains, so it generates the symmetry group given by the classification group of the fermionic invertible phase generated by the Kitaev chain. While the effective field theory for Kitaev chain on an oriented 2d spacetime generates the $\Z_2$ classification of (1+1)D fermionic invertible phase~\cite{Kitaev2001unpaired}, that located on a non-orientable surface instead generates the $\Z_8$ classification~\cite{Fidkowski2010interactions, Kapustin:2014dxa}, since it rather corresponds to classification of (1+1)D fermionic invertible phase with spacetime orientation-reversing  (time-reversal or spatial reflection) symmetry. This implies that the condensation operator acts as a generator of $\Z_8$ group instead of $\Z_2$ in the Hilbert space of $\Z_2$ gauge theory, when the 2d space is non-orientable.

Let us describe the explicit action of this condensation operator on the Hilbert space of the $\Z_2$ gauge theory.
Each state of the $\Z_2$ gauge theory on a 2d space $\Sigma$ can be labeled by the dynamical gauge field for $\Z_2^f$ symmetry, which is identified as Pin$^-$ structure $\eta$ of the surface $\Sigma$. The condensation operator acts on each state by the phase $\exp(\frac{2\pi i}{8}\mathrm{ABK}(\Sigma,\eta))$, which is the partition function for the effective theory of Kitaev chain on the Pin$^-$ surface valued in eighth root of unity~\cite{Kapustin:2014dxa, kobayashi2019pin}.
Here, $\mathrm{ABK}(\Sigma,\eta)$ gives the Arf-Brown-Kervaire invariant of the Pin$^-$ surface. In particular, when the surface $\Sigma$ is given by the real projective plane $\mathbb{RP}^2$, the action on the 2d Hilbert space for $\Z_2$ gauge theory is given by the diagonal matrix $\mathrm{diag}(e^{2\pi i/8},e^{-2\pi i/8})$, which generates $\Z_4$ group up to phase.
This gives a field theoretical explanation for why the dynamics of the honeycomb Floquet code has the enlarged period of dynamics from $\Z_2$ to $\Z_4$ in the presence of a cross-cap.

We note that the condensation defect of the $\psi$ fermion in the honeycomb Floquet code was also described in~\cite{ellison2023floquet}, which realizes the codimension-1 defect of $e\leftrightarrow m$ exchanging symmetry inserted in the 2d space. Our work demonstrates that the measurement period of the honeycomb Floquet code can directly be understood as the condensation of the fermionic particle $\psi$, and derives the action of the dynamics using its expression as the condensation operator.

\section{Discussions}
\label{sec:discussions}
In this work, we described the fault-tolerant logical gates of the stabilizer codes and Floquet codes enabled by putting the code on a non-orientable surface. Here let us describe possible generalizations of the construction of the logical gate considered in this paper.

As described in Sec.~\ref{subsec:condensation}, the action of the $e\leftrightarrow m$ exchanging symmetry on the $\Z_2$ gauge theory on the space $\mathbb{RP}^2$ can be understood via the picture of its ``fermionic dual''; when the $\Z_2$ gauge theory is regarded as a theory obtained by gauging $\Z_2^f$ symmetry of the trivial fermionic invertible phase, the $e\leftrightarrow m$ exchanging symmetry corresponds to evaluating the partition function of the (1+1)D Kitaev chain on the 2d space. Reflecting that the partition function of the Kitaev chain becomes eighth root of unity on $\mathbb{RP}^2$ instead of $\pm 1$, the $e\leftrightarrow m$ exchanging symmetry realized by the dynamics of the honeycomb Floquet code has an enriched action $\mathrm{diag}(e^{2\pi i/8}, e^{-2\pi i/8})$ on the 2d code space.

For future work, it would be very interesting to study the logical gates of $\Z_2$ gauge theory on non-orientable geometry in higher dimensions. For example, (4+1)D $\Z_2$ gauge theory with an emergent fermionic particle supported on a 4d space $\mathbb{RP}^4$ stores a single logical qubit, and has an emergent symmetry that evaluates the partition function of (3+1)D topological superconductor in class DIII~\cite{Metlitski2014} quantized as 16th root of unity~\cite{Kapustin:2014dxa, Tata2022anomalies}. This symmetry is regarded as ``pumping'' a topological superconductor through the 4d space, and acts on the code space by the diagonal matrix $\mathrm{diag}(e^{2\pi i/16}, e^{-2\pi i/16})$, which is the logical $T$ gate up to phase. See also~\cite{fidkowski2023pumping} for a recent development about the logical gate obtained by pumping a fermionic topological phase through the whole space.
It would be interesting to find a realization of this fault-tolerant logical $T$ gate on the (4+1)D $\Z_2$ toric code with a fermionic particle.

\vspace{0.2in}
\noindent{\it Acknowledgements} --- We thank Yu-An Chen, Nat Tantivasadakarn, and Ted Yoder for conversations. GZ also thanks Mohammad Hafezi and Maissam Barkeshli for previous collaboration on ``quantum origami" \cite{Zhu:2017tr}. RK is supported by the JQI postdoctoral fellowship at the University of Maryland. GZ is supported by the U.S. Department of Energy, Office of Science, National Quantum Information Science Research Centers, Co-design Center for Quantum Advantage (C2QA) under contract number DE-SC0012704.

 \bibliography{bibliography, mybib_merge}
 \bibliographystyle{apsrev4-2}

\appendix

\section{Technical details of honeycomb Floquet code}
In this appendix, we give a proof of statements about the honeycomb Floquet code used in the main text.
\subsection{Topological invariance of $\bra{\psi_{r+3}}V(C)\ket{\psi_r}$ }
\label{app:invarianceVC}
Here we show that the expectation value $\bra{\psi_{r+3}}V(C)\ket{\psi_r}$ for $r$ mod 3 is invariant under shifting $C$ by a boundary of any plaquette $p$, i.e.,
\begin{align}
    \bra{\psi_{r+3}}V(C+\partial p)\ket{\psi_r} = \bra{\psi_{r+3}}V(C)\ket{\psi_r}
    \label{eq:topoinvarianceVC}
\end{align}
for any plaquette $p$.
We derive this statement by cases of the color of the plaquette $p$.
Within the code space, the operator $V(C)$ can be expressed as the product of checks on G, B edges as
\begin{align}
V(C)=\left(\prod_{\text{B edges}\in C}v^{(r+2)}_{ij}Z_iZ_j \right)\left(\prod_{\text{G edges}\in C}v^{(r+1)}_{ij}Y_iY_j \right).
\end{align}

\begin{itemize}
    \item When the plaquette $p$ has color R, the operator $V(\partial p)$ is identified as a plaquette stabilizer $w_p W_p$, so we have 
    \begin{align}
        \bra{\psi_{r+3}}V(C)\ket{\psi_r} =\bra{\psi_{r+3}}V(C)V(\partial p)\ket{\psi_r}.
        \end{align}
    Since the product of $ZZ$ checks in $V(\partial p)$ commutes with $YY$ checks on any edges, we have $V(C)V(\partial p) = V(C+\partial p)$. This shows Eq.~\eqref{eq:topoinvarianceVC}.

    \item When the plaquette $p$ has color G, the plaquette stabilizer $w_p W_p$ can be expressed as
    \begin{align}
       w_p W_p = \left(\prod_{\text{R edge}\subset\partial p} v_{ij}^{(r+3)}X_iX_j \right) \left(\prod_{\text{B edge}\subset\partial p} v_{ij}^{(r+2)}Z_iZ_j \right).
        \end{align}
        Since the $XX$ terms of $w_p W_p$ are instantaneous stabilizers at the $(r+3)$-th round, we have
        \begin{align}
        \begin{split}
        \bra{\psi_{r+3}}V(C)\ket{\psi_r} &=    \bra{\psi_{r+3}} w_p W_pV(C)\ket{\psi_r} \\
        &= \bra{\psi_{r+3}}  V(\partial p)V(C)\ket{\psi_r} \\
        &= \bra{\psi_{r+3}}  V(C+\partial p)\ket{\psi_r}.
        \end{split}
        \end{align}
\item When the plaquette $p$ has color B, the plaquette stabilizer $w_p W_p$ can be expressed as
    \begin{align}
       w_p W_p = \left(\prod_{\text{G edge}\subset\partial p} v_{ij}^{(r+1)}Y_iY_j \right) \left(\prod_{\text{R edge}\subset\partial p} v_{ij}^{r}X_iX_j \right).
        \end{align}
        Since the $XX$ terms of $w_p W_p$ are instantaneous stabilizers at the $r$-th round, we have
        \begin{align}
        \begin{split}
        \bra{\psi_{r+3}}V(C)\ket{\psi_r} &=    \bra{\psi_{r+3}} V(C) w_p W_p\ket{\psi_r}\\
        &= \bra{\psi_{r+3}}  V(C)V(\partial p)\ket{\psi_r}\\
        &= \bra{\psi_{r+3}}  V(C+\partial p)\ket{\psi_r}.
        \end{split}
        \end{align}
    \end{itemize}

\subsection{Quadratic property of $V(C)$}
\label{app:quadratic}
Here we show the quadratic property of $V(C)$ described as
\begin{align}
    V(C+C') = (-1)^{\sharp\mathrm{int}(C,C')}  V(C)V(C'),
    \label{eq:quadraticapp}
\end{align}
where $C,C'$ are closed loops and $\sharp\mathrm{int}(C,C')$ is the mod 2 intersection number. This statement also holds for open curves $C, C'$ as long as the ends of $C, C'$ are away from the intersection between $C$ and $C'$. Here, let us restrict ourselves to closed curves for simplicity.
Since one can immediately see that $V(C+C') = \pm V(C)V(C')$ where $\pm$ is an overall phase for the whole Hilbert space, it is sufficient to give a proof of Eq.~\eqref{eq:quadraticapp} within the subspace where $w_pW_p=1$ for all plaquettes. 
For a closed loop $C$, one can easily show that
\begin{align}
    V(C+\partial p) = V(C) w_p W_p \quad \text{for any plaquette $p$}, 
\end{align}
which is valid for the whole Hilbert space. This means that $V(C)$ supported on a closed loop $C$ is topologically invariant $V(C+\partial p) = V(C)$ within the subspace where $w_pW_p=1$. So, due to the topological invariance of $V(C)$, we just have to check Eq.~\eqref{eq:quadraticapp} for a fixed configuration of loops $C,C'$ near the intersection. 

\begin{figure}
    \centering
    \includegraphics[width=0.7\textwidth]{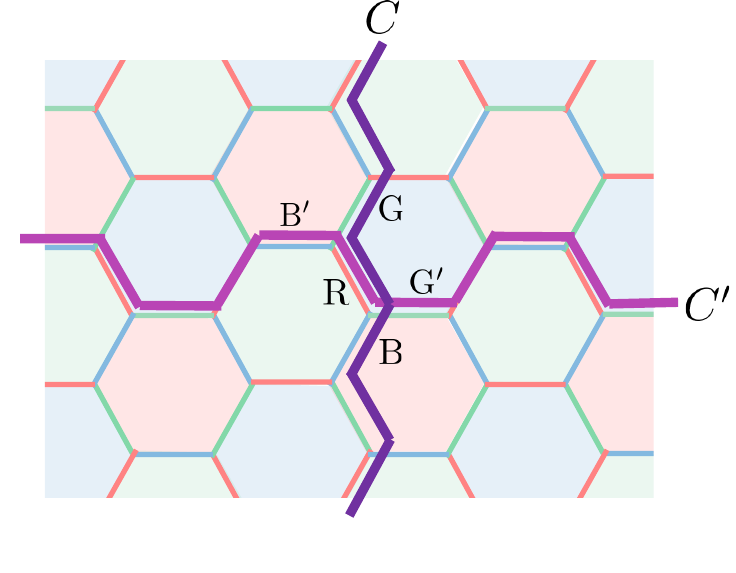}
\caption{The loops $C, C'$ intersect at the single R edge.}
\label{fig:loopoverlap}
\end{figure}

Let us consider the setup where $C$ and $C'$ intersects locally at a single R edge, as described in Fig.~\ref{fig:loopoverlap}.
Then, reordering the terms of $V(C)V(C')$ into $V(C+C')$ near the intersection emits a minus sign
\begin{align}
    \begin{split}
        V(C) V(C') = & (v^{(r+2)}ZZ)_{\mathrm{B}}(v^{(r+1)}YY)_{\mathrm{G}}(v^{r}XX)_{\mathrm{R}} \\
        & \times (v^{(r+2)}ZZ)_{\mathrm{B}'}(v^{(r+1)}YY)_{\mathrm{G}'}(v^{r}XX)_{\mathrm{R}}\\
        =& (v^{(r+2)}ZZ)_{\mathrm{B}}(v^{(r+1)}YY)_{\mathrm{G}} \\
        &\times (v^{(r+2)}ZZ)_{\mathrm{B}'}(v^{(r+1)}YY)_{\mathrm{G}'} \\
        =& -(v^{(r+2)}ZZ)_{\mathrm{B}}(v^{(r+2)}ZZ)_{\mathrm{B}'} \\
        &\times (v^{(r+1)}YY)_{\mathrm{G}}        (v^{(r+1)}YY)_{\mathrm{G}'} \\
        =& -V(C+C'),
        \end{split}
\end{align}
where we omit the checks supported on $C,C'$ irrelevant to the commutation relation between $C$ and $C'$. This implies that the minus sign occurs at each intersection between $C$ and $C'$, which shows the quadratic property Eq.~\eqref{eq:quadraticapp}.

\section{Logical Hadamard gate of $\Z_2$ toric code on a Klein bottle via measurements}
\label{sec:kleinmeasurement}

In this appendix, we describe a way to implement the logical Hadamard gate of the $\Z_2$ toric code on a Klein bottle by a sequence of measurements.
As we discussed in Sec.~\ref{subsec:kleinbottle}, the $\Z_2$ toric code on a Klein bottle admits the logical $\overline{H}_1\otimes\overline{H}_2$ acting on two logical qubits encoded in two cross-caps. We will see that this logical gate can be implemented by a sequence of local measurements, by utilizing a version of Floquet codes called ``Wen plaquette-translation code'' recently defined in Ref.~\cite{aasen2023measurement}. This code realizes the lattice translation of Wen plaquette model by measurements of two-qubit Pauli operators, which corresponds to the $e\leftrightarrow m$ exchanging symmetry of $\Z_2$ toric code. We will see that the Wen plaquette-translation code can be defined on a non-orientable surface as well, and its dynamics realizes the logical Hadamard gate $\overline{H}_1\otimes\overline{H}_2$ on the code space.

\begin{figure*}
    \centering
    \includegraphics[width=1\textwidth]{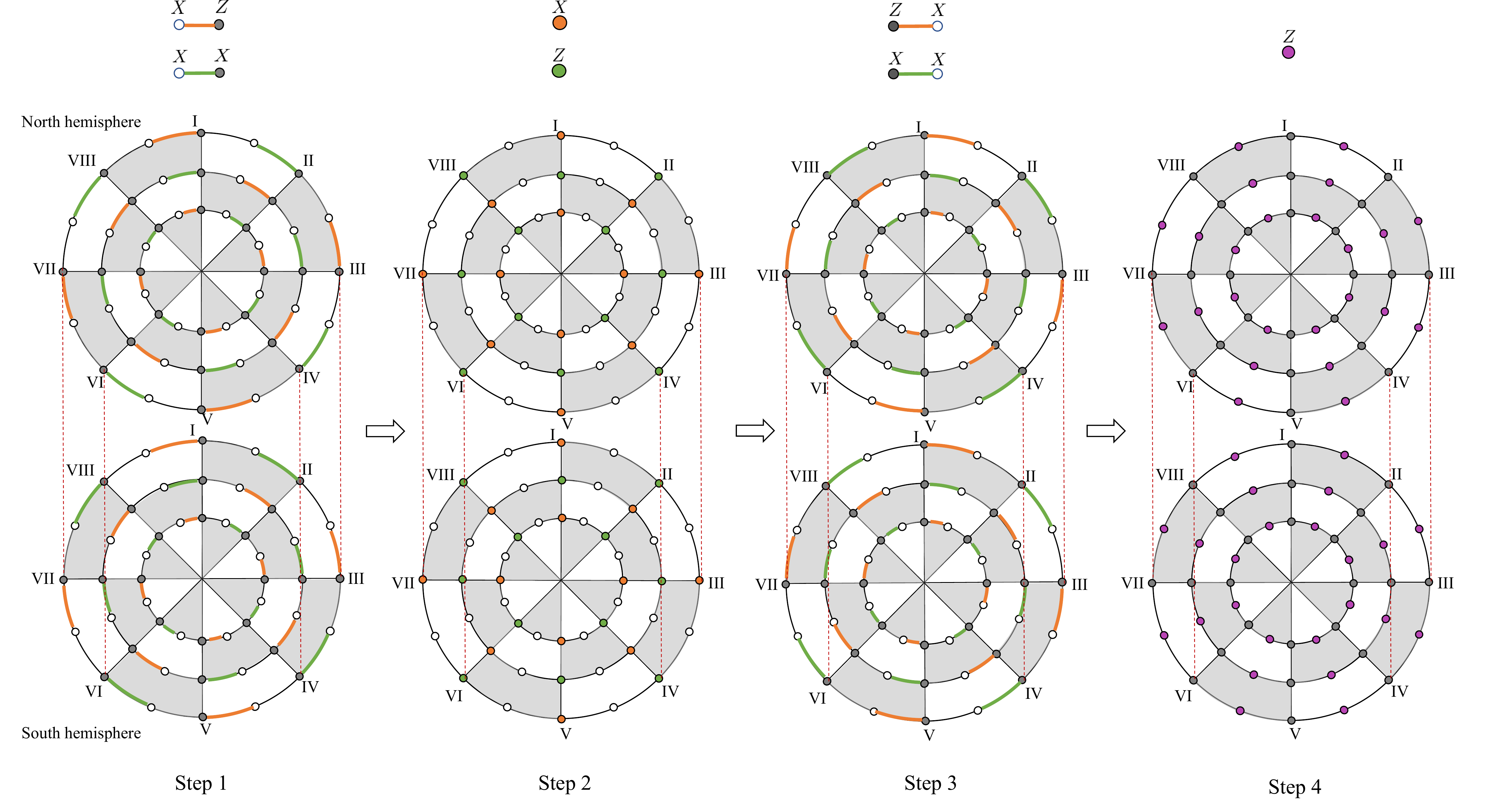}
\caption{The four-step measurements to implement the logical Hadamard gate of the $\Z_2$ toric code on a Klein bottle. }
\label{fig:tc_measurement}
\end{figure*}

Let us consider the toric code on the Klein bottle described in Sec.~\ref{subsec:kleinbottle}. 
We additionally introduce a single qubit in the middle of each edge directed in the horizontal direction, illustrated as white dots in Fig.~\ref{fig:tc_measurement}.

\begin{figure*}
    \centering
    \includegraphics[width=0.55\textwidth]{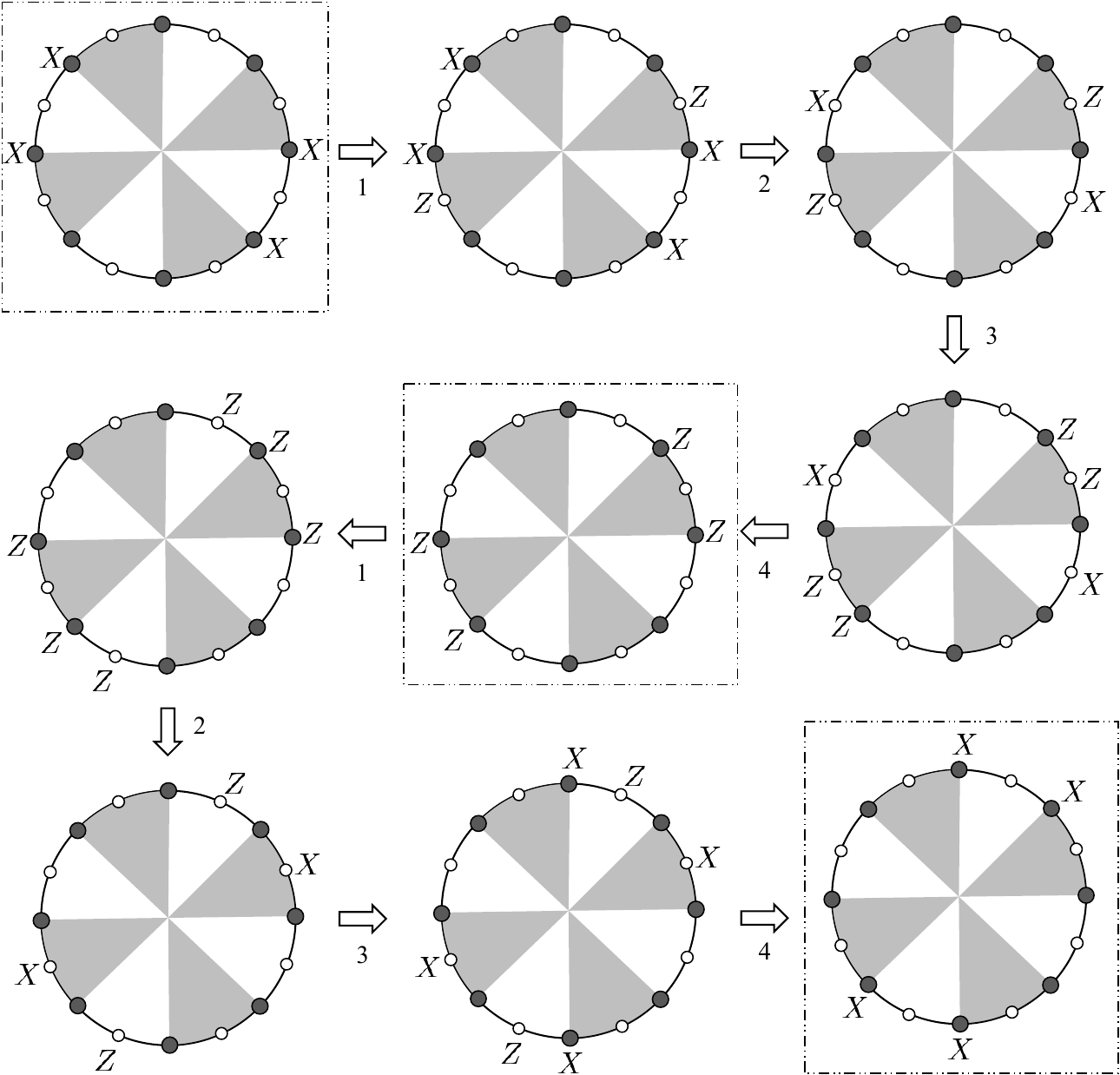}
\caption{The instantaneous stabilizers at the cross-cap of the Klein bottle. One can see that the plaquette stabilizer of the toric code gets translated after four steps.}
\label{fig:crosscap_evolution}
\end{figure*}

Initially, we take the stabilizers as those of the $\Z_2$ toric code on the black qubits and the Pauli $Z$ operator on each auxiliary white qubit.
We then realize the Hadamard gate of this code by the steps of the measurements illustrated in Fig.~\ref{fig:tc_measurement}. One can see that the four steps of measurements induces the translation of the stabilizers, as well as the logical operators. For example, we show the evolution of the instantaneous stabilizer after each step of the measurement in Fig.~\ref{fig:crosscap_evolution}, for the stabilizers at the cross-cap. One can see it implements $e\leftrightarrow m$ exchange of the toric code, which is equivalent to the action of the lattice translation associated with the transversal Hadamard.

\end{document}